\documentclass{article} 
\usepackage{iclr2027_conference,times}

\usepackage{amsmath,amsfonts,bm}

\usepackage{dsfont}

\def\eqref#1{Eq.~(\ref{#1})}

\def\1{\bm{1}}

\DeclareMathAlphabet{\mathsfit}{\encodingdefault}{\sfdefault}{m}{sl}
\SetMathAlphabet{\mathsfit}{bold}{\encodingdefault}{\sfdefault}{bx}{n}

\DeclareMathOperator*{\argmax}{arg\,max}

\usepackage{hyperref}
\usepackage{url}

\usepackage[utf8]{inputenc} 
\usepackage[T1]{fontenc}    
\usepackage{url}            
\usepackage{booktabs}       
\usepackage{amsfonts}       
\usepackage{nicefrac}       
\usepackage{microtype}      
\usepackage{xcolor}         
\usepackage{amssymb}
\usepackage{pifont}

\usepackage{amsmath}
\usepackage{graphicx}
\usepackage{algorithm}
\usepackage{algpseudocode}
\usepackage{pifont}
\usepackage{mathtools}
\usepackage{adjustbox}
\usepackage{enumitem}
\usepackage{wrapfig}
\usepackage[table]{colortbl}
\usepackage{multirow}
\usepackage{makecell}
\usepackage[T1]{fontenc}
\usepackage[utf8]{inputenc}
\usepackage{babel}
\usepackage[font=small]{caption}
\usepackage{pgfplots}
\usepackage{subcaption}
\usepackage{amsmath}
\usepackage{caption}
\usepackage{pgfplotstable}
\usepackage[linewidth=0.5pt]{mdframed}
\usepackage{tikz}
\usepackage{wrapfig}

\usepackage{minitoc}
\usetikzlibrary{pgfplots.groupplots} 

\pgfplotsset{compat=1.18}

\makeatletter
\newcommand{\wb@episource}{}

\makeatother

\definecolor{mydarkblue}{rgb}{0,0.08,0.45}
\definecolor{mydarkred}{rgb}{0.6,0,0}
\definecolor{myblue}{HTML}{268BD2}
\definecolor{mygreen}{HTML}{658354}
\definecolor{check}{rgb}{0,0,0}

\newcounter{todocounter}
\newtoggle{comments}
 \toggletrue{comments} 

\definecolor{DeepGreen}{HTML}{27F542}
\definecolor{softblue}{HTML}{464BB4}

\title{Co-adaptive Rule Routing and Constraining for Structured Knowledge Retrieval}

\author{
\textbf{Zhisheng Qi$^1$, Haoyu Han$^2$, Jiaxi Li$^1$, Bing He$^3$, Huiyuan Chen$^4$,
Franck Dernoncourt$^5$},\\~\textbf{Ryan A. Rossi$^5$, Mahantesh M. Halappanavar$^6$,
Jin Lu$^1$, Yu Zhang$^7$, Tianming Liu$^1$, Yu Wang$^1$
}
\\
\footnotesize
\texttt{
$^1$University of Georgia~~
$^2$Michigan State University
}\\
\texttt{
$^3$Georgia Institute of Technology~~
$^4$Amazon
}
\texttt{
$^5$Adobe Systems
}\\
\texttt{
$^6$PNNL~~
$^7$Texas A\&M University
}}

\usepackage{etoc}
\usepackage{placeins}

\usepackage[most]{tcolorbox}
\usepackage{longtable}
\usepackage{tabularx}
\usepackage{array}
\usepackage{booktabs}
\usepackage{xcolor}

\definecolor{promptback}{RGB}{248,248,248}
\definecolor{promptframe}{RGB}{190,190,190}
\definecolor{rulegray}{RGB}{246,246,246}

\newcolumntype{Y}{>{\raggedright\arraybackslash}X}

\iclrfinalcopy 
\begin{document}

\maketitle
\lhead{Preprint. Under review.}

\maketitle

\begin{abstract}
Structured knowledge retrieval aims to acquire task-relevant evidence from structured knowledge sources for downstream task execution. Existing methods either rely on unconstrained retrieval or predefined query-rule supervision (e.g., author–paper for seeking papers in scientific discovery and anatomy–drug for precision medicine in biomedical analysis), which is rarely available in practice.
To address this limitation, we propose \underline{\textbf{Co}}-Adaptive \underline{\textbf{R}}ule Routing and Constraining R\underline{\textbf{E}}trieval (\textbf{CoRE}), a framework that alternately determines \emph{which rule to route each query to} and \emph{how to retrieve under that rule without requiring predefined query–rule pairs for training supervision}. Specifically, a routing agent assigns each query to a reasoning rule, and the corresponding retriever traverses the structured knowledge base and ranks evidence that satisfies the rule’s constraints.
Since the rule-routing agent and rule-constraining retriever are inherently interdependent, we formulate an alternate optimization that couples reinforcement learning for rule routing with contrastive learning for the rule-constraining retriever. This closed-loop learning enables both components to co-adapt without explicit query–rule supervision.
%
Experiments on STaRK-QA datasets, spanning scientific discovery, biomedical reasoning, and E-commerce enterprise knowledge management domains, verify the superiority of our proposed CoRE. 
Ablations validate co-adapting both router and retriever, and generation-based routing demonstrates cross-dataset transferability and benefits from multi-domain pretraining with target-specific fine-tuning.
Furthermore, marginalizing the routing policy over all queries of each dataset reveals the distribution of rule usage, informing automatic rule refinement and discovery. Our code is available at \href{https://github.com/charlieqi02/Co-Adaptive-Rule-Routing-and-Constraining-Retrieval-}{here}.
\end{abstract}


\vspace{-3ex}
\section{Introduction}
\vspace{-1.5ex}
Retrieval over structured knowledge bases~\citep{nickel2015review} serves as the fundamental component of modern information systems, with tremendous real-world applications such as scientific discovery~\citep{hope2020scisight,auer2020improving}, biomedical reasoning~\citep{bonner2022review}, and enterprise knowledge management~\citep{raina2024question, pipitone2024legalbench}.
Structured knowledge retrieval requires adhering to explicit reasoning constraints, such as retrieving neighbors of specific relation types, traversing predefined meta-paths, or composing symbolic attributes.
For example, a clinician searching for treatment options 
follows $\mathtt{Symptom \rightarrow Diagnosis \rightarrow Treatment}$ to infer appropriate interventions from a patient's observed symptoms and diagnosed condition~\citep{wu2024stark}. A researcher tracing intellectual lineage may traverse $\mathtt{Paper \rightarrow Author \rightarrow Paper}$ to identify related works connected through common authorship~\citep{wu2024stark}. 
A cybersecurity analyst investigating a potential security breach may follow $\mathtt{Alert \rightarrow Host \rightarrow Process \rightarrow Vulnerability}$ to identify the root cause of suspicious activities and assess potential attack vectors~\citep{noel2017bigdata}.
In order to support rule-constrained retrieval across all these domains, a natural question arises:
\vspace{-1ex}
\begin{center}
\textbf{\textit{How can we identify reasoning rules and use them to constrain structured knowledge retrieval?}}
\end{center}
\vspace{-1ex}

\begin{figure}[t!]
    \centering
    \includegraphics[width=\textwidth]{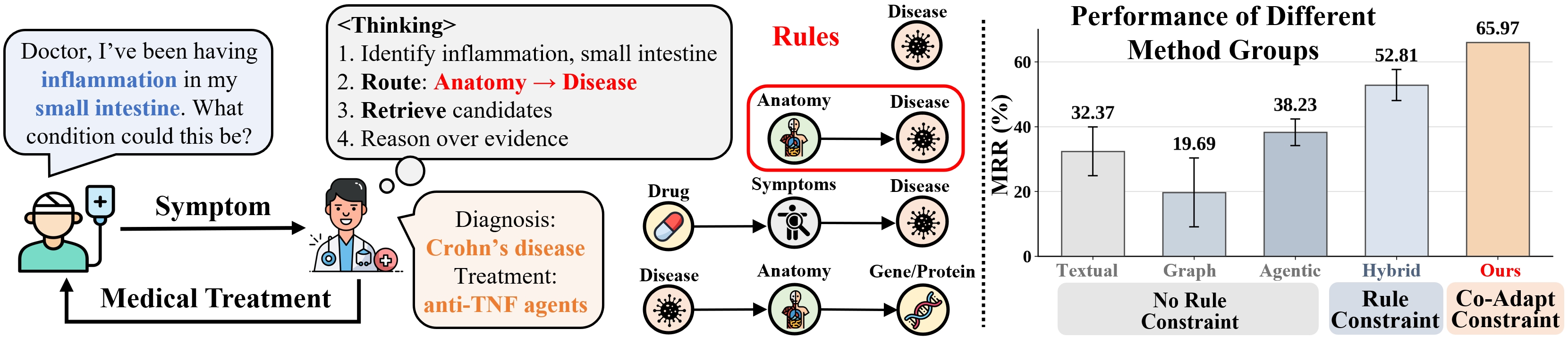}
    \vspace{-3ex}
    \caption{(a) Given a patient’s biomedical query, structured knowledge retrieval is guided by the reasoning rule $\mathtt{Anatomy \rightarrow Disease}$, which a doctor gradually acquires through years of clinical experience. (b) Co-adapting the rule router and rule-constraining retriever further improves retrieval performance.}
    \label{fig:trend}
    \vspace{-3ex}
\end{figure}

Existing structured knowledge retrieval can be broadly categorized into three groups. 
The first group uses structured query and representation methods, including SQL joins, graph traversal, and symbolic rule execution, to identify relevant records, paths, or substructures. Some also learn structure-conditioned embeddings for entity retrieval~\citep{he2021improving,saxena2020improving}. However, they typically use fixed queries or a single scoring pass rather than adaptively identifying and retrieving the next useful evidence.
%
Stepping further, the second group integrates agentic workflows into structured knowledge retrieval, where agents~\citep{han2026reasoning,bai2026graphdancer} dynamically explore the structured knowledge base and select subsequent actions based on the global query and intermediate observations. However, these agentic retrieval methods still struggle to adhere to domain-specific rule constraints.
%
This limitation further motivates the third group, reasoning-constraining retrieval, which leverages predefined reasoning structures, such as meta-paths~\citep{luo2023reasoning,sun2023think,lei2025mixture} and formal logic~\citep{faltings2025enhancing}, to guide structured retrieval. However, the underlying reasoning rule for each query often remains hidden in practice. Existing methods~\citep{luo2023reasoning,han2026reasoning,lei2025mixture} therefore rely on heuristics or LLM knowledge to construct query-rule pair supervision, which introduces construction bias and fails to capture the reasoning rationale required by the query, as empirically demonstrated in Appendix~\ref{app-case-study}

To address the lack of explicit query-rule pairs, we formulate reasoning-rule selection as a latent agentic routing problem, where the routing agent is learned directly from downstream feedback obtained by executing retrieval under the selected rule.
%
This formulation introduces two key challenges.
\textit{First, rule routing and rule-constraining retrieval are inherently coupled.}
The router determines the query-rule assignments used to train rule-constraining retrievers, while retrieval performance provides feedback to improve the routing policy.
\textit{Second, training rule router and rule-constraining retrievers require fundamentally different learning signals.}
The router should identify the rule most likely to yield successful retrieval, whereas each retriever should learn to rank relevant evidence against irrelevant evidence under its assigned rule.
To address these two challenges, we propose \underline{\textbf{Co}}-Adaptive \underline{\textbf{R}}ule Routing and Constraining R\underline{\textbf{E}}trieval (\textbf{CoRE}), an alternate optimization framework that jointly adapts the rule-routing agent and rule-constraining retrievers. Specifically, the rule-routing agent is optimized via reinforcement learning using downstream retrieval feedback, while rule-constraining retrievers are optimized via contrastive learning over queries assigned by the rule-routing agent. Alternating between these two stages enables routing and retrieval to co-adapt, progressively identifying the appropriate reasoning rule for each query and improving retrieval under the corresponding rule constraints. 
Our contributions are summarized as follows:
\vspace{-1.5ex}
\begin{itemize}[leftmargin=*, itemsep=0pt]
    \item \textbf{Reasoning-constraining Retrieval without Assuming Query-Rule Pairs.} We formulate reasoning-rule selection as a latent routing problem and decompose structured retrieval into rule routing and rule-constraining retrieval, which requires no query-rule supervision.

    \item \textbf{Co-Adaptive Rule-Routing Agent and Rule-Constraining Retrievers.} We propose \textbf{CoRE}, an alternate optimization framework where the agentic router assigns each query to an appropriate reasoning rule to adapt the corresponding rule-constraining retriever, while retrieval performance provides feedback to adapt the routing policy.

    \item \textbf{Superior Performance and New Insights.} Experiments show that \textbf{CoRE} achieves the strongest average performance. Ablations validate both adaptation directions, and the token-generation router shows cross-dataset transferability and benefits from multi-domain pretraining. Moreover, the learned routing distributions further provide insights for automatic rule mining and refinement.
\end{itemize}
\vspace{-1.5ex}
\vspace{-2.5ex}
\section{Related Work} \label{sec:rel_works}
\vspace{-2ex}
Our work lies at the intersection of structured knowledge retrieval and constrained agentic reasoning, while additionally addressing the joint adaptation of agent policies and reasoning-constrained retrieval tools. Accordingly, we review related work from three perspectives: (1) Structured knowledge retrieval, (2) Constrained agentic reasoning for retrieval, and (3) Co-adaptation of agents and tools.

\textbf{Structured Knowledge Retrieval.}
Early structured retrieval methods primarily improve how textual and relational signals are combined. MFAR~\citep{li2024mfar} adaptively aggregates retrieval scores across structured fields, while KAR~\citep{xia2025knowledge} augments retrieval with LLM-generated query expansions and relation filtering. More recent methods introduce richer structural reasoning.
MoR~\citep{lei2025mixture} generates a planning graph and combines structural traversal with textual matching, whereas HybGRAG~\citep{lee2025hybgrag} and AvaTaR~\citep{wu2024avatar} use LLM feedback to refine retrieval strategies or tool use.
These methods increasingly adapt how structure is used, but the structural strategy itself is typically generated or optimized separately from the retriever. Our method instead treats rule selection as latent and jointly learns which structural rule to use and how to retrieve under that rule from downstream retrieval feedback.
%


\textbf{Reasoning-Constrained and Agentic Graph Retrieval.}
A complementary line of work moves structural reasoning from retrieval design into the LLM policy itself. RoG~\citep{luo2023reasoning} first generates relation paths as explicit reasoning plans and then retrieves along those paths. ToG~\citep{sun2023think} makes this process interactive by selecting relations and entities during graph exploration. RoE~\citep{han2026reasoning} and GraphDancer~\citep{bai2026graphdancer} further learn such exploration policies through supervised and reinforcement learning.
This progression shifts from predefined retrieval structure to learned traversal decisions, but also requires the LLM to perform low-level graph exploration. Our method separates the two levels: a routing agent selects a schema-valid reasoning rule, while a specialized retriever executes that rule and ranks the reachable evidence. The router is trained directly from retrieval outcomes rather than gold reasoning paths or query--rule labels.

\textbf{Co-Adaptation of Agents and Tools.}
Recent self-improving agents~\citep{jiang2025adaptation} go beyond optimizing a single policy by adapting interacting components in a feedback loop~\citep{huang2026r, li2026online}. CoEvolve~\citep{yang2026coevolve} couples agent learning with an evolving training distribution, WebEvolver~\citep{fang2025webevolver} co-adapts a web agent and world model, and Tool-R0~\citep{acikgoz2026toolr0} jointly improves a task generator and tool-using solver.
Our work brings this co-adaptation principle to structured retrieval. The rule router determines query assignments used for specialized training rule-constrained retrievers, while retriever performance provides the reward for improving the router. 



\section{Notation and Problem Formulation} \label{sec:formu}
In this section, we define key notation and formulate the structured knowledge retrieval problem.

\textbf{Notation.}
Consider a structured knowledge base $\mathcal{G} = (\mathcal{E}, \mathcal{R})$ consisting of an entity set $\mathcal{E}$ and a relation set $\mathcal{R}$, where each relation $r_{ij}\in\mathcal{R}$ connects a pair of entities $e_i,e_j\in\mathcal{E}$. Each entity $e_i$ and relation $r_{ij}$ is associated with its type and descriptive attributes, such as the textual descriptions used in this work.
%
Examples include the abstract of a paper entity or the interaction description of a protein--protein relation.
%
In many domains, effective retrieval follows recurring multi-step reasoning patterns that specify how entities and relations should be progressively traversed and composed. We represent these multi-step reasoning patterns as a set of reasoning constraint rules $\mathcal{K}$
$=\{K|K=(K_1, K_2, \ldots, K_{L_K})\}$ 
where each rule $K$ consists of an ordered sequence of $L_K$ reasoning steps. 
Each step may correspond to an entity type or relation type that guides retrieval toward task-supporting evidence, 
e.g., a scientific literature search requires a reasoning constraint
$\texttt{Author}\!\rightarrow\!\texttt{Paper}\!\rightarrow\!\texttt{Venue}$,
and a biomedical diagnosis requires
$\texttt{Drug}\!\rightarrow\!\texttt{Target}\!\rightarrow\!\texttt{Disease}$.

\textbf{Problem Formulation.}
Given a user query $Q$, the objective is to learn a retrieval model $P_{\boldsymbol{\Theta}}$ parameterized by $\boldsymbol{\Theta}$ that retrieves supporting evidence from the structured knowledge base $\mathcal{G}$. Specifically, each training instance $(Q^{*},S^{*})$ is sampled from the training data distribution $P_{\mathbb{Q}\times\mathbb{S}}^{\text{Train}}$, where $S^{*}\subseteq \mathcal{G}$ denotes the ground-truth evidence for answering $Q^{*}$ and may correspond to a single answer entity, multiple entities, or an entity-relation subgraph. Formally, we aim to learn
\begin{equation}\label{eq-obj}
\boldsymbol{\Theta}^{*}
=
\arg\max_{\boldsymbol{\Theta}}
\mathbb{E}_{(Q^{*},S^{*})\sim P_{\mathbb{Q}\times\mathbb{S}}^{\text{Train}}}
[
P_{\boldsymbol{\Theta}}
(
S^{*}\mid Q^{*},\mathcal{G}
)].
\end{equation}
Unlike prior structure retrieval methods~\cite{han2026reasoning,lei2025mixture}, we do not assume the availability of query--rule pair
supervision or access (i.e., joint distribution $P_{\mathbb{Q}\times\mathbb{K}}$). Instead, we assume only the availability of query--evidence pairs sampled from $P_{\mathbb{Q}\times\mathbb{S}}^{\text{Train}}$, where $\mathbb{S}$ denotes the ground-truth supporting evidence provided by the retrieval dataset. The reasoning rule that connects each query to its supporting evidence remains latent and should be inferred automatically.

\begin{figure}[t!]
    \centering
    \includegraphics[width=\textwidth]{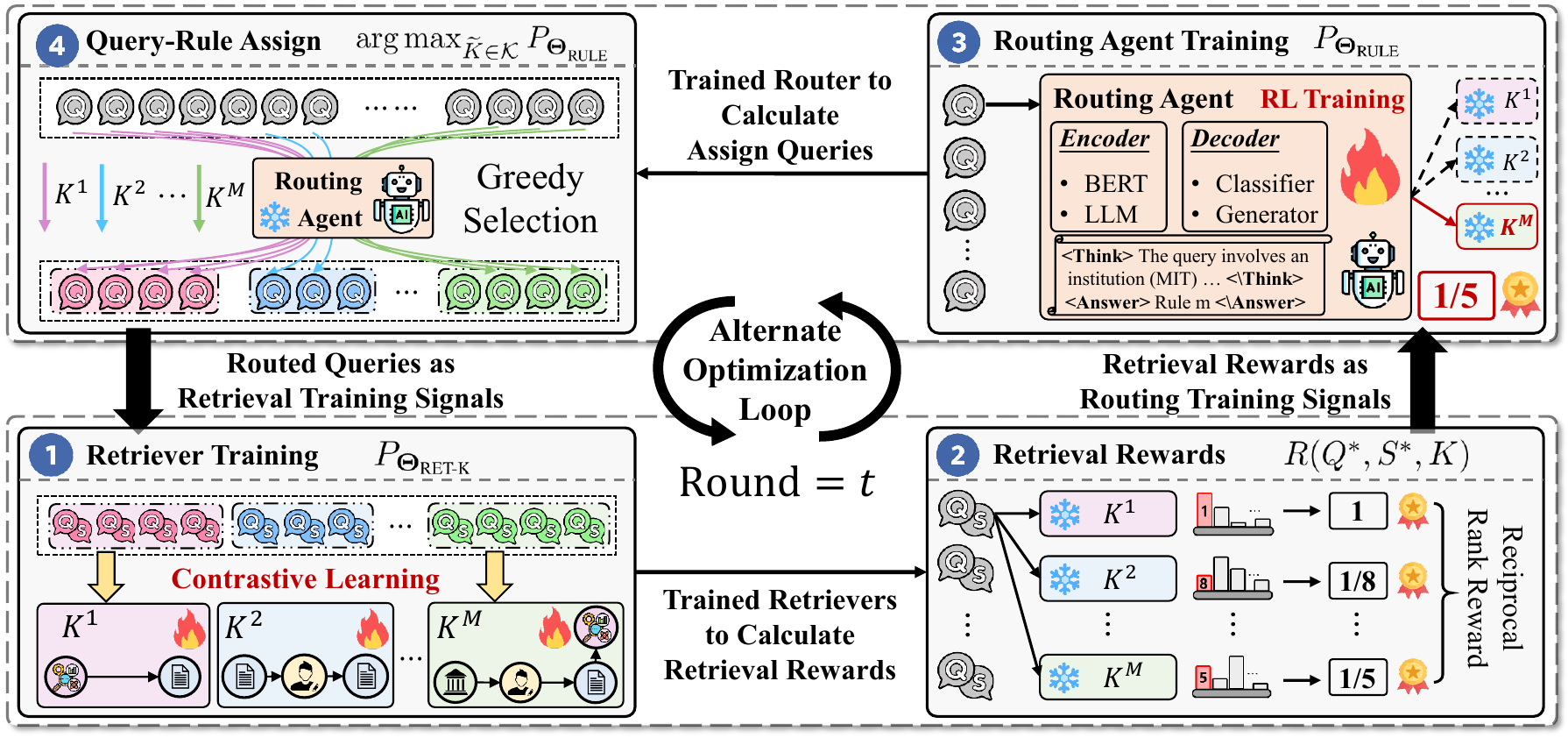}
    \vspace{-3ex}
    \caption{Overview of the CoRE framework. CoRE alternates between optimizing rule-constraining retrievers and the rule-routing agent. Given the current query--rule assignments, each retriever is first trained on queries assigned to its corresponding rule (\ding{172}). The resulting retrieval performance is converted into rule-level reward signals (\ding{173}), which are used to update the routing policy (\ding{174}). The updated router then reassigns queries to rules (\ding{175}), providing new supervision for the next round of retriever optimization. This alternating process enables the routing policy and rule-constraining retrievers to co-adapt through a closed feedback loop.
    }
    \label{fig-framework}
    \vspace{-2ex}
\end{figure}


\section{Framework}\label{sec: Framework}
\vspace{-2ex}
In this section, we first provide an overview of CoRE and then detail the design of the rule-routing agent module in Section~\ref{sec:Router} and the rule-constraining retrieval module in Section~\ref{sec:Retriever}, respectively. 
Next, we present the alternative optimization together with the corresponding training strategies for both components in Section~\ref{sec: optimization}. Finally, we describe the inference procedure in Section~\ref{sec: inference}.

As discussed earlier, a key challenge in rule-constraining structured knowledge retrieval is the absence of query--rule pair supervision, making it impossible to directly train a rule-routing policy through supervised learning. Instead, the available training data contains only observable query--evidence pairs $(Q,S)$. Consequently, directly optimizing the retrieval objective
$P_{\boldsymbol{\Theta}}(S \mid Q,\mathcal{G})$ in \eqref{eq-obj}
requires jointly inferring the latent reasoning rule
$P_{\boldsymbol{\Theta}_{\text{RULE}}}(K \mid Q,\mathcal{K})$
and retrieving evidence conditioned on the inferred rule,
$P_{\boldsymbol{\Theta}_{\text{RET-K}}}(S \mid Q,K,\mathcal{G})$.
To address this challenge, we model the reasoning rule as a latent variable and optimize the factorized retrieval likelihood as:
\begin{equation}
\resizebox{0.93\linewidth}{!}{$\displaystyle
\{\boldsymbol{\Theta}_{\text{RET-K}}^*\}_{K=1}^{|\mathcal{K}|},
\boldsymbol{\Theta}_{\text{RULE}}^*
=
\argmax_{\{\boldsymbol{\Theta}_{\text{RET-K}}\}_{K=1}^{|\mathcal{K}|},
\boldsymbol{\Theta}_{\text{RULE}}}
\mathbb{E}_{(Q^*, S^*)\sim P^{\text{Train}}_{\mathbb{Q}\times \mathbb{S}}}
\sum_{K\in\mathcal{K}}
\underbrace{
P_{\boldsymbol{\Theta}_{\text{RET-K}}}(S\mid Q,K,\mathcal{G})
}_{\text{Rule-constraining Retriever}}
\times
\underbrace{
P_{\boldsymbol{\Theta}_{\text{RULE}}}(K\mid Q,\mathcal{K})
}_{\text{Rule-routing Agent}}
$}
\label{eq:framework_decomposition}
\end{equation}
%
%
where $P_{\boldsymbol{\Theta}_{\text{RULE}}}(K\mid Q,\mathcal{K})$ denotes the rule routing policy that estimates the probability of assigning query $Q$ to reasoning rule $K$ from the rule space $\mathcal{K}$, and $P_{\boldsymbol{\Theta}_{\text{RET-K}}}(S\mid Q,K,\mathcal{G})$ denotes the rule-constrained retriever that estimates the probability of retrieving evidence $S$ from the structured knowledge source $\mathcal{G}$ conditioned on rule $K$. 

Intuitively, the rule-routing policy acts as a high-level planner that selects the most suitable latent rule for each query, while each rule-constraining retriever specializes in retrieving evidence under its corresponding structural constraint. As shown in Figure~\ref{fig-framework}, CoRE couples them through a four-stage loop. Starting from uniform 
query--rule assignments, \ding{172} each retriever is trained on the queries assigned to its rule; \ding{173} the trained retrievers execute the candidate rules, and the rank of the ground-truth evidence is used to construct routing rewards; \ding{174} these rewards optimize the routing agent toward rules that produce better retrieval; and \ding{175} the updated agent reassigns queries, providing new supervision for the next round of retriever training.
%


Importantly, the rule-routing policy and rule-constraining retrievers are intrinsically interdependent. The rule-routing policy determines the query--rule assignments that supervise retriever training, whereas retriever performance provides reward signals for refining the rule-routing policy. This mutual dependency precludes optimizing either component in isolation and motivates the alternative optimization framework presented in Section~\ref{sec: optimization}.

\subsection{Agentic Rule Routing Module}\label{sec:Router}
\vspace{-1ex}
The rule-routing module aims to infer the latent reasoning rule that best explains a query. Formally, given a query $Q$ and a candidate rule set
$\mathcal{K}$,
the rule-routing module models the rule assignment distribution
$P_{\boldsymbol{\Theta}_{\text{RULE}}}(K\mid Q,\mathcal{K})$.
To accommodate both discriminative rule classification and generative rule prediction, we adopt a general encoder--decoder formulation:
$
P_{\boldsymbol{\Theta}_{\text{RULE}}}(K\mid Q,\mathcal{K})
=
f_{\text{Decoder}}\!\left(
f_{\text{Encoder}}(Q),\mathcal{K}
\right).
$
The encoder transforms the query into a latent representation
$\mathbf{h}_Q\in\mathbb{R}^{d}$, and the decoder maps
$\mathbf{h}_Q$ into a rule-selection decision.

\textbf{Encoder.}
We consider two encoder instantiations:
\begin{equation}
\mathbf{h}_Q
=
f_{\text{Encoder}}(Q)
=
\left\{
\begin{array}{ll}
\mathrm{BERT}(Q)_{\texttt{[CLS]}},
&
\text{BERT-based encoder},
\\[2mm]
\mathrm{LLM}(Q)_{\text{Last}},
&
\text{LLM-based encoder}.
\end{array}
\right.
\end{equation}

The BERT-based encoder extracts the final hidden state of the \texttt{[CLS]} token to capture sentence-level semantic information. In contrast, the LLM-based encoder uses the final-layer hidden state of the last token, enabling richer contextual representations that better capture complex reasoning patterns.

\vspace{-1ex}
\textbf{Decoder.}
Given the query representation $\mathbf{h}_Q$, the decoder defines the rule-selection distribution as:
\vspace{-0.2ex}
\begin{equation}
f_{\text{Decoder}}(\mathbf{h}_Q,\mathcal{K})
=
\left\{
\begin{array}{ll}
\mathrm{Softmax}
\!\left(
\mathbf{W}\mathbf{h}_Q+\mathbf{b}
\right),
&
\text{Classifier-based decoder},
\\[2mm]
\prod_{t=1}^{T}
P(y_t \mid y_{<t},Q,\mathcal{K}),
&
\text{Generation-based decoder}.
\end{array}
\right.
\end{equation}

For the classifier-based decoder,
$\mathbf{W}\in\mathbb{R}^{|\mathcal{K}|\times d}$ and
$\mathbf{b}\in\mathbb{R}^{|\mathcal{K}|}$ are learnable parameters that produce a categorical distribution over the candidate rules in $\mathcal{K}$. A rule is subsequently selected through argmax selection or categorical sampling. Alternatively, the generation-based decoder formulates rule selection as a conditional generation task, where the generated sequence
$Y=\{y_t\}_{t=1}^{T}$ is parsed into a rule assignment. The prompting format is provided in Appendix~\ref{app:prompts}.

Combining the encoder and decoder choices yields three variants of CoRE:
\textbf{BERT-C}, which combines a BERT-based encoder with a classifier-based decoder;
\textbf{LLM-C}, which combines an LLM-based encoder with a classifier-based decoder; and
\textbf{LLM-G}, which combines an LLM-based encoder with a generation-based decoder.
We do not instantiate a BERT-based generation variant, as coupling BERT with a generative decoder would introduce additional architecture design complexity and training choices beyond the scope of this study.
\vspace{-1ex}
\subsection{Rule-constraining Retrieval Module}\label{sec:Retriever}
\vspace{-1ex}
Given a query $Q$ and a reasoning rule $K$ selected by the rule-routing agent, the corresponding rule-constraining retriever identifies evidence that is both semantically relevant to the query and structurally consistent with the selected rule. Formally, the likelihood of retrieving candidate evidence $S_i$ for query $Q$ under reasoning rule $K$ is defined as
$
P_{\boldsymbol{\Theta}_{\text{RET-K}}}(S_i \mid Q,K,\mathcal{G})=
\frac{
\exp (
\Phi(Q,S_i,K)
)
}{
\sum_{S_j\in\mathcal{C}(Q,K)}
\exp (
\Phi(Q,S_j,K)
)
},
$
where $\mathcal{C}(Q,K)$ is the candidate evidence obtained by constrained depth-first-search (DFS) under rule $K$, and $\Phi(Q,S_i,K)$ is a rule-conditioned relevance score measuring the compatibility between query $Q$, candidate evidence $S_i$, and rule $K$. $P_{\boldsymbol{\Theta}_{\text{RET-K}}}$ provides a general formulation of rule-constrained retrieval. We instantiate the framework for knowledge graph retrieval, as experimented in Section~\ref{sec-experiment}.

For rule-constrained knowledge graph retrieval, the key is to assign each candidate entity a relevance score that reflects both its compatibility with the query and its selected rule. We therefore define the rule-conditioned relevance score $\Phi(Q,S_i,K)$ by combining rule-constrained structural information with query-dependent relevance signals.
To incorporate the rule constraint, a rule $K$ specifies a sequence of entity types, such as \textit{Anatomy $\rightarrow$ Disease $\rightarrow$ Drug}. The retriever performs constrained graph traversal following this sequence, which restricts the search space to entities reachable under $K$. For each resulting candidate $S_i$, let $(v_{i,1},\ldots,v_{i,L})$ denote the traversal path leading to it, where $v_{i,L}=S_i$.
To incorporate query relevance, we measure the BM25 similarity between the query and the textual description of each entity along this path:
$
\mathbf{x}_i
=
\left[
\operatorname{BM25}(Q,v_{i,1}),
\ldots,
\operatorname{BM25}(Q,v_{i,L})
\right].
$
The resulting vector $\mathbf{x}_i$ therefore summarizes how relevant the entire rule-constrained path is to the query.
Finally, we combine the query-dependent path features with the representation of the selected rule. Specifically, $\mathbf{x}_i$ and $f_{\text{embed}}(K)$ are provided as inputs to a learnable scoring function:
$
\Phi(Q,S_i,K)
=
f_{\text{score}}
\left(
\mathbf{x}_i,
f_{\text{embed}}(K)
\right),
$
where $f_{\text{embed}}(K)$ represents the selected rule embedding and $f_{\text{score}}$ is parameterized by $\boldsymbol{\Theta}_{\text{RET-}K}$. In this way, $\Phi(Q,S_i,K)$ jointly captures query relevance along the constrained traversal path and the semantics of the selected rule.
\newpage
\subsection{Alternative Optimization}\label{sec: optimization}
\vspace{-1ex}
Since the rule routing policy determines the latent query–rule assignments that supervise rule-constraining retrievers, while retrieval performance in turn provides feedback for improving rule routing, we therefore formulate their co-adaptation as a bi-level optimization problem and optimize them through an alternate optimization strategy~\citep{bai2026alternating,tarzanagh2024online} that iteratively updates the rule-constraining retrievers and rule routing policy.
\begin{equation}
\label{eq:alternating_optimization}
\boldsymbol{\Theta}^{t+1} = \left\{
\begin{aligned}
\boldsymbol{\Theta}_{\text{RET-K}}^{t+1}
&=
\argmax_{\boldsymbol{\Theta}_{\text{RET-K}}}
\mathcal{L}_{\text{RET}}
(\{\boldsymbol{\Theta}_{\text{RET-}K^m}^{t}\}_{m=1}^{|\mathcal{K}|}, \boldsymbol{\Theta}_{\text{RULE}}^{t}
),
~\ ~
\text{with}~\boldsymbol{\Theta}_{\text{RULE}}^{t}~\text{fixed},
\\
\boldsymbol{\Theta}_{\text{RULE}}^{t+1}
&=
\argmax_{\boldsymbol{\Theta}_{\text{RULE}}}
\mathcal{L}_{\text{RULE}}(
\boldsymbol{\Theta}_{\text{RULE}}^{t},
\{\boldsymbol{\Theta}_{\text{RET-}K^m}^{t+1}\}_{m=1}^{|\mathcal{K}|}),
\
\text{with } \{\boldsymbol{\Theta}_{\text{RET-}K^m}^{t+1}\}_{m=1}^{|\mathcal{K}|}~\text{fixed}.
\end{aligned}
\right.
\end{equation}
where $\mathcal{L}_{\text{RET}}$ and $\mathcal{L}_{\text{RULE}}$ denote the optimization objectives for the rule-constraining retrievers and the rule routing policy, respectively. The specific formulations are introduced next.

%

\textbf{Rule-constraining Retrieval Optimization.}
With the rule routing policy fixed, each rule-constrained retriever is trained on the queries assigned to its rule by that policy.
Let
$
\argmax_{\widetilde{K}\in\mathcal{K}}
P_{\boldsymbol{\Theta}_{\text{RULE}}^t}(\widetilde{K}\mid Q^*)
$
denote the rule assigned to training pair $(Q^*,S^*)$. 
Therefore, given a fixed rule assignment, the marginalization over all candidate rules in~\eqref{eq:framework_decomposition} reduces to the assigned rule, yielding the following:
\begin{equation}
\label{eq:retriever_lower_objective}
\mathcal{L}_{\text{RET}} = \mathbb{E}_{(Q^{*},S^{*})\sim P^{\text{Train}}_{\mathbb{Q}\times \mathbb{S}}}\sum_{K\in\mathcal{K}}\mathds{1}(K=\argmax_{\widetilde{K}\in\mathcal{K}}\{P_{\boldsymbol{\Theta}_{\text{RULE}}^{t}}(\widetilde{K}|Q^*)\})P_{\boldsymbol{\Theta}_{\text{RET-K}}^{t}}(S^*|Q^*, K, \mathcal{G})
\end{equation}
where $P_{\boldsymbol{\Theta}_{\text{RULE}}^{t}}(\widetilde{K}|Q^*)$ and $P_{\boldsymbol{\Theta}_{\text{RET-K}}^{t}}(S^*|Q^*, K, \mathcal{G})$ are calculated following Sections~\ref{sec:Router}-\ref{sec:Retriever}.

\textbf{Rule routing Optimization.} 
With the rule-constrained retrievers fixed, we next optimize the rule routing policy using retrieval feedback. Intuitively, the policy should assign high probability to reasoning rules that lead to successful evidence retrieval and low probability to rules that result in poor retrieval performance. For each training query $Q^*$, the rule routing policy samples a reasoning rule $K \sim P_{\boldsymbol{\Theta}_{\text{RULE}}^{t}}(\cdot \mid Q^*)$ which is subsequently evaluated by the optimized rule-constrained retrievers:
\begin{equation}
\label{eq:router_upper_objective}
\begin{aligned}
\mathcal{L}_{\text{RULE}}
=
\mathbb{E}_{(Q^{*},S^{*})\sim P^{\text{Train}}_{\mathbb{Q}\times \mathbb{S}},~K\sim P_{\boldsymbol{\Theta}^t_{\text{RULE}}}(\cdot|Q^{*})}
\ell_{\text{Rank}}(Q^*, S^*, K, \{\boldsymbol{\Theta}_{\text{RET-K}}^{t+1}\}_{K\in\mathcal{K}}).
\end{aligned}
\end{equation}
The ranking objective $\ell_{\text{Rank}}$ is defined based on the retrieval rank of the ground-truth evidence. Given a training pair $(Q^*,S^*)$ and a reasoning rule $K$, the corresponding rule-constrained retriever ranks all candidate evidence in $\mathcal{C}(Q^*,K)$ as in Section~\ref{sec:Retriever}. Let $\mathrm{rank}(S^* \mid Q^*,K)$ denote the ranking position of the ground-truth evidence $S^*$ among these candidates. We convert this ranking position into a scalar retrieval reward using Reciprocal Rank:
$
R(Q^*,S^*,K)
=
\frac{1}
{\mathrm{rank}(S^* \mid Q^*,K)}
$.
Depending on the policy parameterization, $\ell_{\text{Rank}}$ is optimized differently: classifier-based policies rank over all candidate rules, whereas generation-based policies optimize over sampled rollouts.


\begin{itemize}[leftmargin=*, itemsep=0pt]

\item \textbf{Classifier-based Rule Routing.}
For classifier-based policies, the model predicts a distribution over all candidate rules, allowing rewards to be evaluated for every $K\in\mathcal{K}$. We compute a rule-wise advantage
$
A(Q^*,S^*,K)
=
R(Q^*,S^*,K)
-
\frac{1}{|\mathcal{K}|}
\sum_{K'\in\mathcal{K}}
R(Q^*,S^*,K')
$,
and optimize the expected advantage under the policy:
$
\ell_{\text{Rank}}
=
\sum_{K\in\mathcal{K}}
P_{\Theta_{\text{RULE}}}(K\mid Q^*)
A(Q^*,S^*,K).
$

\item \textbf{Generation-based Rule Routing.}
For generation-based policies, evaluating all candidate rules is computationally prohibitive. Instead, we sample a rollout set
$
\{K_1,\ldots,K_N\}
\sim
P_{\Theta_{\text{RULE}}}(\cdot\mid Q^*)
$
and compute group-relative advantages from their Reciprocal Rank-based retrieval rewards. Following GRPO, the policy is optimized using
$
\ell_{\text{Rank}}
=
\frac{1}{N}
\sum_{i=1}^{N}
\min
(
r_iA_i,
\mathrm{clip}(r_i,1-\epsilon,1+\epsilon)A_i
),
$
where $r_i$ is the policy ratio and $A_i$ is the group-relative advantage.

\end{itemize}
 
\vspace{-1ex}
\subsection{Inference}\label{sec: inference}
\vspace{-1ex}
At inference time, given a query $Q$, the optimized rule routing policy first identifies the most probable reasoning rule. Conditioned on the selected rule, the corresponding rule-constrained retriever constructs a candidate evidence set $\mathcal{C}(Q,K^*)$, assigns each candidate a rule-conditioned relevance score $\Phi(Q,S_i,K^*)$, and returns the highest-ranked evidence:
\begin{equation}
K^{*}
=
\argmax_{K\in\mathcal{K}}
P_{\boldsymbol{\Theta}_{\mathrm{RULE}}}(K\mid Q)
\qquad
\hat{S}
=
\operatorname{TopK}_{S_i\in\mathcal{C}(Q,K^*)}
\Phi(Q,S_i,K^*).
\end{equation}
The retrieved evidence set $\hat{S}$ is subsequently provided to downstream task execution.


\newpage

\begin{table*}[t!]
\centering
\footnotesize
\caption{Retrieval performance on AMAZON, MAG, and PRIME, comparing baselines with \textsc{CoRE-BERT-C}, \textsc{CoRE-LLM-C}, and \textsc{CoRE-LLM-G}. Best and second-best results are \textbf{bold} and \underline{underlined}, respectively.}
\vspace{-2ex}
\label{tab-merged}
\resizebox{\linewidth}{!}{
\setlength\tabcolsep{4pt}
\begin{tabular}{ll|cccc|cccc|cccc|cccc}
\toprule
\multirow{3}{*}{\textbf{Category}} & \multirow{3}{*}{\makecell[l]{\textbf{Retrieval} \\\textbf{Baseline}}}  & \multicolumn{4}{c|}{\textbf{AMAZON}} & \multicolumn{4}{c|}{\textbf{MAG}} & \multicolumn{4}{c|}{\textbf{PRIME}} & \multicolumn{4}{c}{\textbf{AVERAGE}} \\
\cmidrule(lr){3-6} \cmidrule(lr){7-10} \cmidrule(lr){11-14} \cmidrule(lr){15-18}
& & H@1 & H@5 & R@20 & MRR & H@1 & H@5 & R@20 & MRR & H@1 & H@5 & R@20 & MRR & H@1 & H@5 & R@20 & MRR \\
\midrule
\multirow{4}{*}{\textbf{Textual}}
    & BM25       & 40.62 & 62.18 & 50.89 & 50.59 & 23.11 & 44.47 & 44.02 & 33.23 & 11.96 & 25.67 & 28.31 & 18.35 & 25.23 & 44.11 & 41.07 & 34.06\\
    & Ada        & 39.16 & 62.73 & 53.29 & 50.35 & 29.08 & 49.61 & 48.36 & 38.62 & 12.63 & 31.49 & 36.00 & 21.41 & 26.96 & 47.94 & 45.88 & 36.79\\
    & Multi-Ada  & 40.07 & 64.98 & 55.12 & 51.55 & 25.92 & 50.43 & 50.80 & 36.94 & 15.10 & 33.56 & 38.05 & 23.49 & 27.03 & 49.66 & 47.99 & 37.33\\
    & DPR        & 15.29 & 47.93 & 44.49 & 30.20 & 10.51 & 35.23 & 42.11 & 21.34 &  4.46 & 21.85 & 30.13 & 12.38 & 10.09 & 35.00 & 38.91 & 21.31\\
\midrule
\multirow{2}{*}{\textbf{Graph}}
    & QAGNN      & 26.56 & 50.01 & 52.05 & 37.75 & 12.88 & 39.01 & 46.97 & 29.12 &  8.85 & 21.35 & 29.63 & 14.73 & 16.10 & 36.79 & 42.88 & 27.20\\
    & ToG        & --    & --    & --    & --    & 13.16 & 16.17 & 11.30 & 14.18 &  6.07 & 15.71 & 13.07 & 10.17 &  9.62 & 15.94 & 12.18 & 12.18\\
\midrule
\multirow{5}{*}{\textbf{Hybrid}}
    & AvaTaR     & 49.87 & 69.16 & \underline{60.57} & 58.70 & 44.36 & 59.66 & 50.63 & 51.15 & 18.44 & 36.73 & 39.31 & 26.73 & 37.56 & 55.18 & 50.17 & 45.53\\
    & KAR        & \textbf{54.20} & 68.70 & 57.24 & 61.29 & 50.47 & 69.57 & 60.28 & 58.65 & 30.35 & 49.30 & 50.81 & 39.22 & 45.01 & 62.52 & 56.11 & 53.05\\
    & MFAR$^{*}$ & 41.20 & 70.00 & 58.50 & 54.20 & 49.00 & 69.60 & 71.70 & 58.20 & \underline{40.90} & \underline{62.80} & \underline{68.30} & \underline{51.20} & 43.70 & 67.47 & \underline{66.17} & 54.53\\
    & HYBGRAG    & --    & --    & --    & --    & \textbf{65.40} & 75.31 & 65.70 & \underline{69.80} & 28.56 & 41.38 & 43.58 & 34.49 & 46.98 & 58.35 & 54.64 & 52.15\\
    & MoR        & \underline{52.19} & \underline{74.65} & 59.92 & \underline{62.24} & 58.19 & \underline{78.34} & \underline{75.01} & 67.14 & 36.41 & 60.01 & 63.48 & 46.92 & \underline{48.93} & \underline{71.00} & 66.14 & \underline{58.77}\\
\midrule
\multirow{3}{*}{\textbf{Agentic}}
    & RoG        & 40.93 & 62.79 & 36.96 & 49.25 & 24.52 & 45.33 & 30.93 & 35.03 & 14.87 & 22.41 & 23.28 & 19.36 & 26.77 & 43.51 & 30.39 & 34.55 \\
    & RoE        & 42.90 & 64.17 & 37.42 & 50.96 & 27.18 & 46.62 & 31.41 & 37.76 & 14.18 & 33.80 & 37.93 & 23.63 & 28.09 & 48.20 & 35.59 & 37.45 \\
    & GraphDancer & 46.27 & 67.43 & 65.91 & 55.18 & 29.33 & 64.67 & 66.97 & 44.36 & 18.67 & 29.33 & 44.53 & 28.54 & 31.42 & 53.81 & 59.14 & 42.69 \\
\midrule
\multirow{3}{*}{\shortstack{\textbf{CoRE}\\\textbf{(Ours)}}}
    & \cellcolor{gray!30} \makebox[2em][c]{BERT}\makebox[1.5em][c]{--}\makebox[1em][l]{C}
    & \cellcolor{gray!30} 50.71 & \cellcolor{gray!30} \textbf{76.19} & \cellcolor{gray!30} \textbf{61.39} & \cellcolor{gray!30} \textbf{62.46} 
    & \cellcolor{gray!30} \underline{63.37} & \cellcolor{gray!30} \textbf{87.78} & \cellcolor{gray!30} \textbf{83.51} & \cellcolor{gray!30} \textbf{74.17}
    & \cellcolor{gray!30} \textbf{49.34} & \cellcolor{gray!30} \textbf{76.40} & \cellcolor{gray!30} \textbf{80.81} & \cellcolor{gray!30} \textbf{61.27}
    & \cellcolor{gray!30}\textbf{54.47} & \cellcolor{gray!30}\textbf{80.12} & \cellcolor{gray!30}\textbf{75.24} & \cellcolor{gray!30}\textbf{65.97}\\
    & \cellcolor{gray!30} \makebox[2em][c]{LLM}\makebox[1.5em][c]{--}\makebox[1em][l]{C}
    & \cellcolor{gray!30} 50.83 & \cellcolor{gray!30} 75.55 & \cellcolor{gray!30} 60.85 & \cellcolor{gray!30} 62.21
    & \cellcolor{gray!30} 60.24 & \cellcolor{gray!30} 83.03 & \cellcolor{gray!30} 77.64 & \cellcolor{gray!30} 71.53
    & \cellcolor{gray!30} 45.97 & \cellcolor{gray!30} 74.12 & \cellcolor{gray!30} 79.18 & \cellcolor{gray!30} 58.57
    & \cellcolor{gray!30} 52.35 & \cellcolor{gray!30} 77.57 & \cellcolor{gray!30} 72.56 & \cellcolor{gray!30} 64.10 \\
    & \cellcolor{gray!30} \makebox[2em][c]{LLM}\makebox[1.5em][c]{--}\makebox[1em][l]{G}
    & \cellcolor{gray!30} 50.26 & \cellcolor{gray!30} 73.24 & \cellcolor{gray!30} 60.92 & \cellcolor{gray!30} 58.64
    & \cellcolor{gray!30} 59.14 & \cellcolor{gray!30} 82.39 & \cellcolor{gray!30} 77.92 & \cellcolor{gray!30} 69.25
    & \cellcolor{gray!30} 35.54 & \cellcolor{gray!30} 55.81 & \cellcolor{gray!30} 59.01 & \cellcolor{gray!30} 44.94
    & \cellcolor{gray!30} 48.31 & \cellcolor{gray!30} 70.48 & \cellcolor{gray!30} 65.95 & \cellcolor{gray!30} 57.61 \\
\bottomrule
\end{tabular}
}
\vspace{-3ex}
\end{table*}
\section{Experiment}\label{sec-experiment}
\vspace{-2ex}
This section aims to experimentally verify the effectiveness of CoRE. We first introduce experimental settings and then conduct experiments to answer the following questions: 
\begin{itemize}[leftmargin=*, nosep]
    \item \S~\ref{sec:main_results} - $\mathbf{Q}_1$: \textit{How does CoRE compare with existing structured retrieval methods?}
    \item \S~\ref{sec:ablation} - $\mathbf{Q}_2$: \textit{How do router adaptation and retriever adaptation affect the performance?}
    \item \S~\ref{sec:trans} - $\mathbf{Q}_3$: \textit{Can the token-generation routing agents transfer learned routing knowledge across datasets and benefit from multi-domain pretraining?}
    \item \S~\ref{sec:dynamics} - $\mathbf{Q}_4$: \textit{How do the routing agent and retrievers co-adapt across alternate rounds?}
\end{itemize}

\vspace{-1ex}
\subsection{Experimental Setup}\label{sec:setup}
\vspace{-1ex}
\textbf{Datasets and evaluation.}
We evaluate on three STaRK-QA~\citep{wu2024stark} datasets: AMAZON, MAG, and PRIME.
We define 6, 8, and 65 rules by sampling all ordered entity-type sequences connected by dataset-schema-permitted relations in the traversal direction.
%
Following our motivation, training uses only query--evidence pairs, while query-rule assignments remain latent and are inferred by the routing agent. We report Hits@1, Hits@5, Recall@20, and MRR.

\vspace{-1ex}
\textbf{Model variants.}
Following Section~\ref{sec:Router}, we study three routing architectures: CoRE-BERT-C, which uses a BERT encoder and classification head; CoRE-LLM-C, which uses an LLM encoder and classification head; and CoRE-LLM-G, which generates rule selections as tokens. All variants use the same rule-constraining retriever design and undergo three alternate optimization rounds. The generation-based router is initialized from Qwen2.5-3B-Instruct
and trained with eight sampled rollouts per query on four NVIDIA H200 GPUs.

\vspace{-1ex}
\textbf{Baselines.}
We compare against textual retrievers, including BM25~\citep{robertson2009probabilistic}, Ada~\citep{oai}, Multi-Ada~\citep{oai}, and DPR~\citep{dpr}; graph-based methods, including QAGNN~\citep{yasunaga2021qa} and ToG~\citep{sun2023think}; hybrid methods, including AvaTaR~\citep{wu2024avatar}, KAR~\citep{xia2025knowledge}, MFAR~\citep{li2024mfar}, HYBGRAG~\citep{lee2025hybgrag}, and MoR~\citep{lei2025mixture}; and agentic graph methods, including RoG~\citep{luo2023reasoning}, RoE~\citep{han2026reasoning}, and GraphDancer~\citep{bai2026graphdancer}.

Further details, including model implementation and hyperparameter settings, are in Appendix~\ref{app:experiment-details}.
\vspace{-1ex}


\subsection{$\mathbf{Q}_1$-Main Results}\label{sec:main_results}

\vspace{-1ex}
Table~\ref{tab-merged} shows that CoRE-BERT-C achieves the strongest average performance. 
\vspace{-1ex}
\begin{itemize}[leftmargin=*, itemsep=0pt]
    \item \textbf{Compared with textual and graph-based methods}, all CoRE variants outperform the textual and graph-based baselines. These gains support using structural rule constraints for retrieval, particularly on MAG and PRIME, where textual matching alone performs poorly.

    \item \textbf{Compared with hybrid methods}, CoRE-BERT-C outperforms MoR, achieving the highest average performance among evaluated hybrid methods. While MoR relies on LLM-generated constraints that may not align with downstream retrieval needs, CoRE learns rule routing directly from retrieval feedback. This advantage is more pronounced on PRIME, where its larger domain-specific rule space makes it harder for LLMs to select appropriate rules based on their general knowledge alone.
    
    \item \textbf{Compared with agentic graph methods}, all CoRE variants achieve higher average scores than RoG, RoE, and GraphDancer. 
    One explanation is that agentic methods rank evidence through step-by-step graph exploration based on intermediate node observations, whereas CoRE first constrains the search space with a reasoning rule and then ranks the fetched evidence using a rule retriever, which is more effective in these text-rich graphs.
   
    \item \textbf{Compared among CoRE variants}, CoRE-BERT-C achieves the strongest average performance, followed by CoRE-LLM-C and CoRE-LLM-G. With the same classification head, BERT's stronger results suggest that its sentence-level representation is better suited to rule routing in this setting. Both classification-based routers outperform generation-based routing on average, likely because they use reward feedback from all candidate rules rather than sampled rollouts. Limited rollouts provide less reward feedback and higher-variance gradient estimates, reducing training efficiency~\citep{cheng2020trajectory}. This limitation is most shown on PRIME, where eight rollouts per query cover only a small portion of its 65-rule space. Generation-based routing nevertheless supports cross-dataset transfer and multi-domain foundational pretraining, as in Section~\ref{sec:trans}.
\end{itemize}

\vspace{-3ex}
\subsection{$\mathbf{Q}_2$-Ablation Study of Co-Adapting Router and Retriever} \label{sec:ablation}
\begin{table}[t!]
\centering
\footnotesize
\caption{Ablation of routing agent and retriever adaptation.
\textbf{w/o Agent-Adapt} fixes the router to a pre-trained or zero-shot model while continuing retriever training.
\textbf{w/o Retriever-Adapt} replaces learned rule-constraining retrievers with fixed BM25-traversal scoring while continuing routing agent training.
Best results are in \textbf{bold}.}
\vspace{-2ex}
\label{tab:ablation}
\setlength\tabcolsep{5pt}
\resizebox{\linewidth}{!}{%
\begin{tabular}{l|cc|cc|cc|cc}
\toprule
\multirow{2}{*}{\textbf{Method}}
    & \multicolumn{2}{c|}{\textbf{AMAZON}}
    & \multicolumn{2}{c|}{\textbf{MAG}}
    & \multicolumn{2}{c|}{\textbf{PRIME}}
    & \multicolumn{2}{c}{\textbf{Average}} \\
    & H@1 & MRR & H@1 & MRR & H@1 & MRR & H@1 & MRR \\
\midrule
\rowcolor{gray!20}
CoRE-BERT-C
    & \textbf{50.71} & \textbf{62.46} & \textbf{63.37} & \textbf{74.17} & \textbf{49.34} & \textbf{61.27} & \textbf{54.47} & \textbf{65.97} \\
w/o Agent-Adapt (unadapted BERT)
    & 14.63 & 19.83 & 28.23 & 35.05 & 5.15 & 8.15 & 16.00 & 21.01 \\
w/o Agent-Adapt (GPT-5-mini)
    & 26.70 & 32.72 & 28.23 & 31.87 & 4.14 & 6.66 & 19.69 & 23.75 \\
w/o Retriever-Adapt (BM25-traversal)
    & 49.87 & 61.45 & 55.51 & 65.45 & 38.86 & 49.70 & 48.08 & 58.87 \\

\bottomrule
\end{tabular}
}
\vspace{-4ex}
\end{table}

\vspace{-1ex}
Since CoRE co-adapts rule routing and rule-constrained retrieval, we ablate each adaptation direction in Table~\ref{tab:ablation} and further examine their joint benefit in Appendix~\ref{sec:extended-ablation} due to space limitations.

\textbf{Without routing agent adaptation.} Replacing the learned routing agent with a fixed pre-trained BERT or zero-shot GPT-5-mini policy substantially reduces performance across all three datasets, despite continued retriever training. This suggests that poorly aligned query-rule assignments limit both rule routing quality and rule-constraining retriever specialization.

\textbf{Without retriever adaptation.}
We replace the learned rule-constraining retrievers with fixed BM25-traversal scoring, ranking candidates under the selected rule by the mean query-node BM25 similarity along their traversal paths. This preserves the rule-constraining candidate space and BM25 relevance signals while removing learnable rule-conditioned scoring. The routing agent is still trained on the resulting retrieval rewards.
Performance drops most on PRIME, followed by MAG and AMAZON. This aligns with the rule distributions in Figure~\ref{fig-rule-dom}: PRIME contains more diverse multi-hop rules that benefit from learning how to combine relevance signals across traversal steps, whereas the less-node-hop rules common on MAG and AMAZON rely largely on a few BM25 scores.

\vspace{-2ex}
\subsection{$\mathbf{Q}_3$-Transferability of Generation-based Rule Routing} \label{sec:trans}
\vspace{-1ex}

\begin{wrapfigure}{r}{0.35\textwidth}
    \vspace{-3.5ex}
    \centering
    \includegraphics[width=\linewidth]{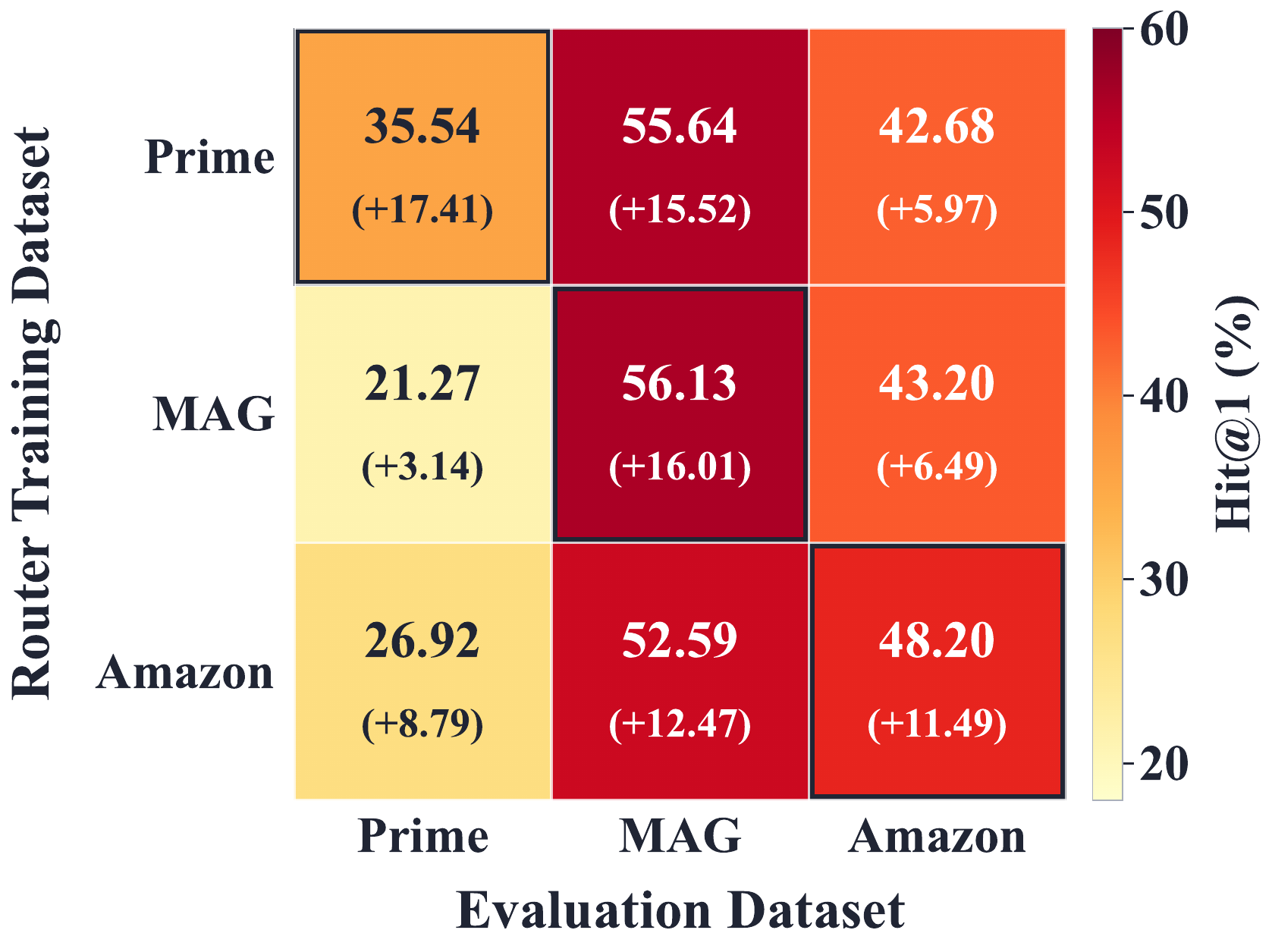}
    \vspace{-4ex}
    \caption {Cross-dataset H@1 of CoRE-LLM-G. Parentheses show gains over zero-shot Qwen with trained retrievers.
    }
    \label{fig:cross-eval}
    \vspace{-4ex}
\end{wrapfigure}

Unlike fixed classification heads, generation-based routers leverage LLMs to generate rules in the token space, enabling natural transfer to unseen rule sets across datasets. We therefore examine whether CoRE-LLM-G transfers across datasets and further benefits from multi-domain pretraining.

\textbf{Cross-dataset transfer.}
We independently train CoRE routers on AMAZON, MAG, and PRIME and evaluate each router across all three datasets. We report H@1 together with its improvement over an unadapted Qwen router, using the same trained retrievers for a controlled comparison. During cross-dataset transfer, only task description, graph schema, and candidate rules are replaced; the router remains fixed and is paired with target-domain retrievers. Figure~\ref{fig:cross-eval} shows that all transferred routers outperform zero-shot Qwen, suggesting that retrieval-feedback training learns transferable mappings between query semantics and reasoning patterns rather than memorizing dataset-specific query-rule mappings.

\begin{wrapfigure}{r}{0.35\textwidth}
    \vspace{-1ex}
    \centering
    \includegraphics[width=\linewidth]{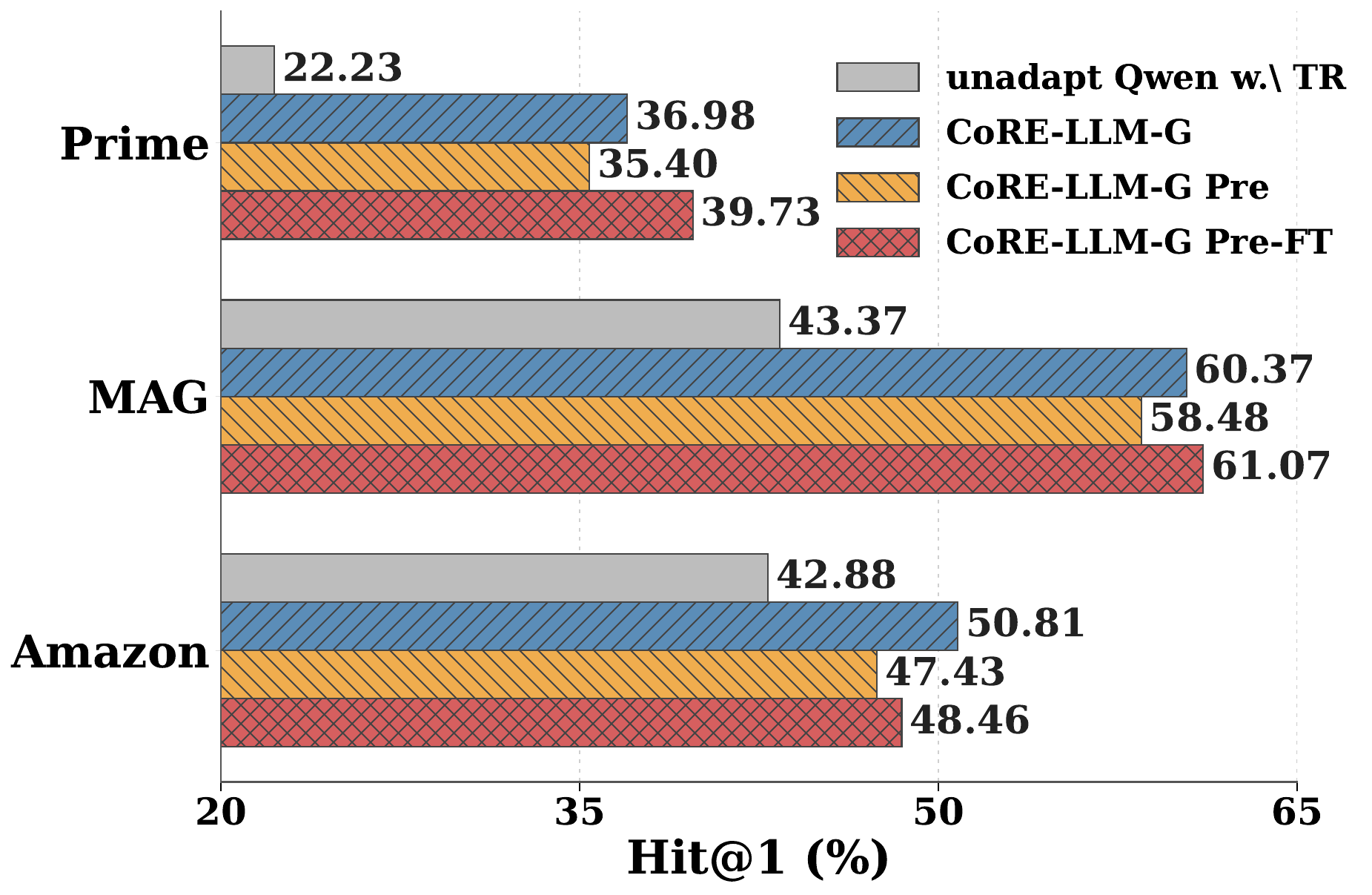}
    \vspace{-4ex}
    \caption{H@1 of CoRE-LLM-G under different training using Qwen2.5-32B-Instruct as the base router. Multi-domain pretraining followed by target-specific fine-tuning performs best.}
    \label{fig:foundation}
    \vspace{-2ex}
\end{wrapfigure}
\textbf{Foundation routing agent.}
Motivated by the cross-dataset transfer results above, we next ask whether routing experience from different domains can be combined to build a stronger shared router. We use Qwen2.5-32B-Instruct in this study to provide greater model capacity for consolidating heterogeneous routing knowledge across multiple domains. We compare four settings: unadapt Qwen w.\ TR uses the base Qwen model without routing training; CoRE-LLM-G trains a separate router on each dataset; CoRE-LLM-G Pre jointly trains one router for one epoch on the combined AMAZON, MAG, and PRIME training sets and directly evaluates it on each dataset; Finally, CoRE-LLM-G Pre-FT further fine-tunes this shared router on each target dataset. All four settings use the same Qwen2.5-32B-Instruct backbone and the same trained retrievers for the corresponding evaluation dataset.

As shown in Figure~\ref{fig:foundation}, the jointly pretrained routing CoRE-LLM-G Pre consistently outperforms unadapt Qwen w.\ TR, indicating that retrieval feedback from different domains provides useful shared knowledge for rule selection.
Its performance remains slightly below dataset-specific training CoRE-LLM-G, suggesting that each domain still has its own routing patterns that are difficult to capture through shared pretraining alone.
However, once further fine-tuned on the target dataset, the CoRE-LLM-G Pre-FT achieves the best generation-based routing performance overall. These results suggest a natural \emph{pretrain-then-specialize} paradigm: first learn general routing knowledge across domains, and then adapt it to the reasoning patterns of each target domain.

\begin{wrapfigure}{r}{0.5\textwidth}
    \vspace{-5ex}
    \centering
    \includegraphics[width=\linewidth]{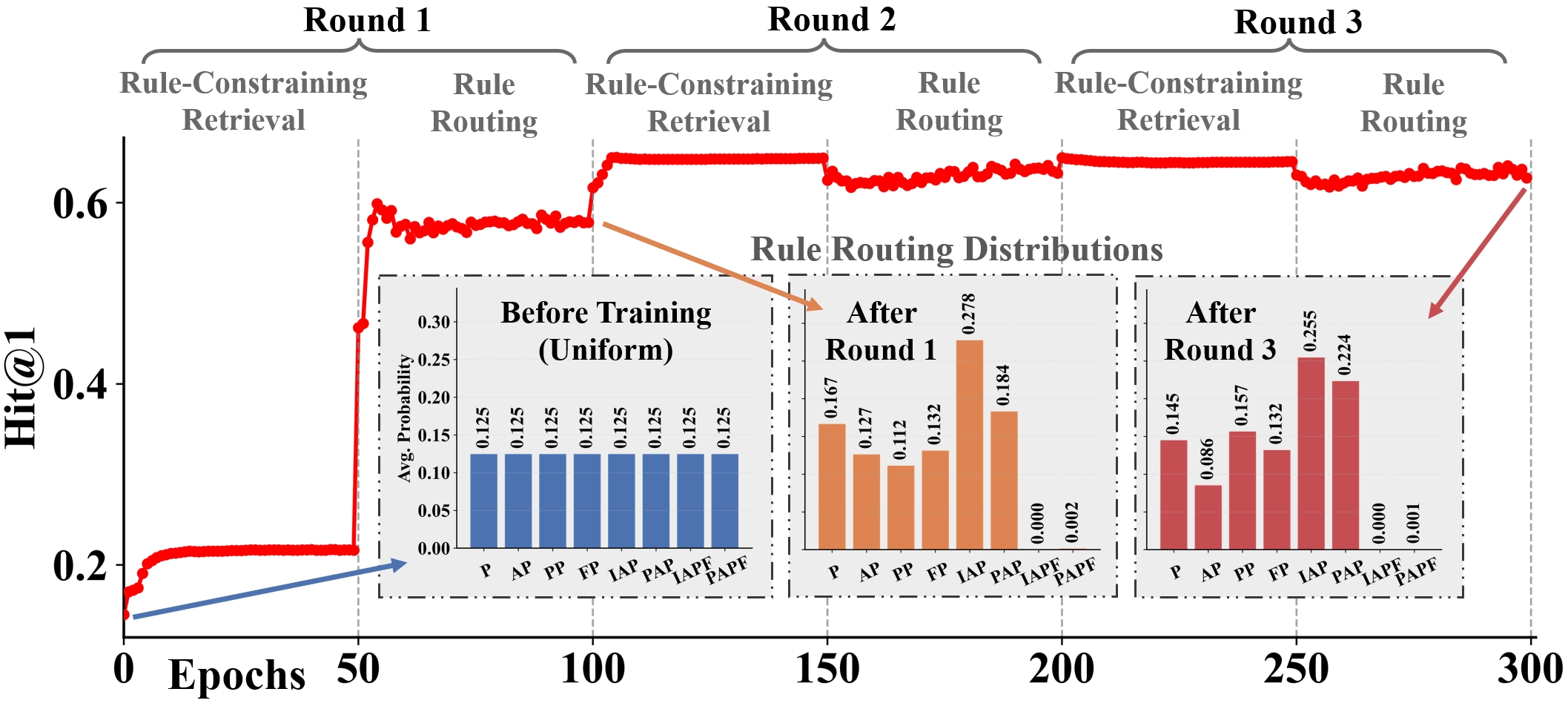}
    \vspace{-3ex}
    \caption{Optimization dynamics of CoRE-BERT-C on MAG. Each round alternates between retriever and router adaptation; the insets show the routing distributions before training and after Rounds 1 and 3.}
    \label{fig:trend}
    \vspace{-2ex}
\end{wrapfigure}

\vspace{-2ex}
\subsection{$\mathbf{Q}_4$-Alternate Optimization Dynamics}
\label{sec:dynamics}
\vspace{-1ex}
Figure~\ref{fig:trend} shows clear co-adaptation across three rounds: router training sharply improves H@1 from about $21\%$ to $58\%$, and the resulting assignments further boost the next retriever update above $62\%$. Later rounds provide smaller refinements as the two components approach convergence.
The inset distributions show how the router differentiates candidate rules without query-rule supervision. Starting from a uniform distribution, after the Round 1 rule-routing training, the agent assigns the highest probability to \texttt{IAP} (institution$\rightarrow$author$\rightarrow$paper) and \texttt{PAP} (paper$\rightarrow$author$\rightarrow$paper), while nearly eliminating the longer \texttt{IAPF} (institution$\rightarrow$author$\rightarrow$paper$\rightarrow$field of study) and \texttt{PAPF} (paper$\rightarrow$author$\rightarrow$paper$\rightarrow$field of study) rules.
By Round 3, \texttt{IAP} remains dominant, while more probability shifts toward \texttt{PAP} and \texttt{PP} (paper$\rightarrow$paper). This indicates that the router identifies useful institution-, authorship-, and citation-based patterns and refines their use as the retrievers improve. This can also support a future direction toward automatic rule discovery as new datasets continually emerge, where high-reward rules guide rule expansion while consistently unused rules are pruned.

A case study demonstrating that general LLMs struggle to identify the appropriate rules without explicit domain rule supervision is also provided in the Appendix~\ref{app-case-study} due to space limitations.
%







\vspace{-2ex}
\section{Conclusion}
\label{sec:conlude}
\vspace{-2ex}
In this work, we study a novel problem of structured knowledge retrieval under rule-based constraints. We introduce CoRE, a structured knowledge retrieval framework that jointly learns rule routing and rule-constraining retrieval without explicit query–rule supervision. CoRE treats rule selection as a latent decision and employs alternating-based bi-level optimization to jointly specialize rule-constraining retrievers and optimize the routing agent through retrieval feedback.
Experiments demonstrate that CoRE consistently outperforms existing baselines, while ablation studies confirm the complementary importance of routing-agent optimization and rule-constraining retriever adaptation.
Analysis of the optimization dynamics further reveals how routing and retrieval progressively co-adapt through iterative feedback.
These findings also motivate automatic rule refinement, where high-reward routing patterns guide the generation and adjustment of reasoning rules.

\newpage

\subsection*{AI use statement}
In this work, we used generative AI tools to assist with manuscript review and language polishing. In particular, we leveraged generative AI to provide feedback following the ICLR review rubric, including identifying potential strengths and weaknesses, and used this feedback to further revise and improve the manuscript. We also used generative AI to assist with writing Python code for figure generation and visualization. We have carefully reviewed all AI-assisted content and take full responsibility for the final manuscript, including all text, claims, code, figures, and other artifacts produced with the aid of generative AI.

\subsection*{Reproducibility}
We provide the code and datasets necessary to reproduce the results shown in Figures~\ref{fig:cross-eval}, \ref{fig:foundation}, \ref{fig:trend}, and Tables~\ref{tab-merged}, \ref{tab:ablation} in the GitHub repository: \href{https://github.com/charlieqi02/Co-Adaptive-Rule-Routing-and-Constraining-Retrieval-}{link}. The repository contains the full CoRE implementation, data preprocessing pipelines, training and evaluation scripts, experiment configurations, hyperparameters, and scripts for reproducing the reported tables and figures.



\bibliography{reference}
\bibliographystyle{iclr2027_conference}

\newpage
\appendix

\begingroup

\refstepcounter{part}
\addcontentsline{toc}{part}{Appendix}

{\huge\bfseries Appendix}

\vspace{2em}

{\LARGE\bfseries Table of Contents}

\vspace{0.4em}
\hrule
\vspace{0.5em}

\etocsetnexttocdepth{subsection}

\etocsettocstyle{}{}

\etocsetstyle{section}
{}
{}
{\noindent
  \makebox[2.5em][l]{\bfseries\etocnumber}
  \bfseries\etocname
  \nobreak\leaders\hbox{\normalfont.\kern0.4em}\hfill
  \bfseries\etocpage
  \par
  \vspace{0.45em}
}
{}

\etocsetstyle{subsection}
{}
{}
{\noindent
  \hspace*{2.5em}
  \makebox[3.8em][l]{\etocnumber}
  \etocname
  \nobreak\leaders\hbox{.\kern0.4em}\hfill
  \etocpage
  \par
  \vspace{0.15em}
}
{}

\localtableofcontents

\vspace{0.5em}
\hrule

\endgroup

\clearpage


\section{Experiment Setting Details}
\label{app:experiment-details}

We provide additional implementation details for the rule-constraining retrievers and the three routing variants used in CoRE: \textsc{BERT-C}, \textsc{LLM-C}, and \textsc{LLM-G}. Unless otherwise specified, all variants use the same rule sets, retriever architecture, candidate construction, and evaluation protocol.

\textbf{Rule Space and Candidate Retrieval.}
We construct rule sets from the schema of each STaRK-QA dataset, resulting in 6 rules for AMAZON, 8 for MAG, and 65 for PRIME. For each query--rule pair, graph traversal under the corresponding rule produces a candidate pool, from which up to 200 candidate entities are retained for ranking. Each rule is represented by a 384-dimensional embedding. At inference time, the router selects one rule, and only the corresponding rule-constraining retriever is used to rank candidates within the resulting structural search space.

\textbf{Rule-constraining Retrievers.}
Each rule is associated with a lightweight retriever that scores candidates using the path-level relevance features introduced in Section~\ref{sec: Framework}. The candidate feature vector and rule embedding are independently projected into 256-dimensional hidden representations and jointly mapped to a scalar relevance score. Retriever training uses pairwise ranking with one positive candidate and up to 200 negative candidates per query. We optimize the retrievers with AdamW using a learning rate of $5\times10^{-5}$, weight decay $0.01$, and batch size 512. Each retriever is trained for at most 50 epochs, and the checkpoint with the best validation Hit@1 is retained.

The downstream supervision for router optimization is derived from rank of ground-truth evidence under each rule-constraining retriever. For query $q$ and rule $k$, we define routing reward as reciprocal rank, $R(q,k)=1/\operatorname{rank}(q,k)$, such that higher reward corresponds to better retrieval under selected rules. For queries with multiple correct entities, we use the best rank among correct answers.

\textbf{\textsc{BERT-C} Router.}
The classification-based BERT router uses \texttt{bert-base}. Queries are tokenized to a maximum length of 128 tokens, and the final \texttt{[CLS]} representation is passed to a dataset-specific linear classification head over the candidate rules. Both the encoder and classification head are fine-tuned jointly. The router is trained with AdamW using batch size 256, learning rate $1\times10^{-5}$, and weight decay $0.01$. For each query, the training target is the rule with the highest downstream retrieval reward, i.e., $y_q=\arg\max_k R(q,k)$.

\textbf{\textsc{LLM-C} Router.}
The LLM classification router uses Qwen2.5-3B as the query encoder. Inputs are truncated to 256 tokens, and the final hidden state of the last non-padding token is passed to a linear rule-classification head. We adapt the LLM backbone using LoRA on the query, key, value, and output projection layers while training the classification head directly. We use a learning rate $1\times10^{-5}$ for LoRA parameters, with LoRA rank 32 and scaling factor 64 for AMAZON and MAG; the selected PRIME configuration uses rank 16 and scaling factor 32. Training is performed in BF16.

Unlike \textsc{BERT-C}, which uses only the highest-reward rule as a hard target, \textsc{LLM-C} exploits the complete rule-wise reward vector. Let $\bar{R}_q$ denote the mean reward across candidate rules for query $q$. The router is optimized using
\[
\mathcal{L}_{\mathrm{LLM-C}}
=
-\frac{1}{B}\sum_q\sum_k
\log p_\theta(k\mid q)\bigl(R(q,k)-\bar{R}_q\bigr)
-\gamma H[p_\theta(\cdot\mid q)].
\]
Thus, rules with above-average retrieval rewards are promoted, while below-average rules are suppressed. The entropy regularization term encourages exploration over candidate rules during training. At inference time, the rule with the highest predicted probability is selected.

\textbf{\textsc{LLM-G} Router.}
The generation-based router uses \texttt{Qwen 2.5 3B} and formulates rule selection as conditional generation. The input contains the task instruction, natural-language query, and textual descriptions of all candidate rules. The model generates a reasoning trace followed by a single rule identifier in the format \texttt{<think> reasoning </think><answer> rule </answer>}. The model is adapted with LoRA on the query, key, value, and output projections, using rank 32, scaling factor 64, and dropout 0.05.

The router is optimized with group-relative policy optimization (GRPO). For each query, we sample eight rule-selection trajectories and evaluate the selected rules using their corresponding frozen retrievers. The reward combines format correctness and downstream retrieval quality as $R(q,y)=0.2R_{\mathrm{format}}(y)+R(q,k(y))$, where $k(y)$ is the rule extracted from the generated response. Invalid rule identifiers receive no retrieval reward. We use BF16 training, a global prompt batch size of 32, eight generations per prompt, and three training epochs. The KL coefficient is $0.01$, and generation uses temperature $0.9$ during training. At inference time, decoding is deterministic and the selected rule is extracted from the \texttt{<answer>} field.

For the verified PRIME configuration, the generation-based router is additionally warm-started with supervised fine-tuning on filtered teacher demonstrations before GRPO. Demonstrations are retained only when they satisfy the required output format and select a rule whose retriever ranks a correct entity near the top of the candidate list. This warm-up primarily teaches the required generation format and provides an initial routing policy before optimization with downstream retrieval feedback.

\textbf{Bi-Level Training.}
All CoRE variants are optimized for three bi-level rounds. Within each round, the routing policy is fixed while the rule-constraining retrievers are updated using the current query--rule assignments. The retrievers are then frozen, and their retrieval ranks provide the feedback used to update the routing policy. The updated router subsequently reassigns queries to rules for the next round. Both router and retriever parameters are warm-started from the preceding round rather than reinitialized, allowing routing and retrieval to progressively co-adapt.

\textbf{Evaluation.}
For each query, the learned router selects a single reasoning rule, after which the corresponding rule-constraining retriever ranks the candidates reachable under that rule. We report Hit@1, Hit@5, Recall@20, and mean reciprocal rank (MRR). Classification-based routers select the highest-probability rule, while the generation-based router uses the rule identifier produced by deterministic greedy decoding.

\textbf{Computational Resources.}
Experiments were conducted on a server equipped with four NVIDIA H200 NVL GPUs. Generative SFT and GRPO training use all four GPUs with distributed data parallelism, whereas rule-constraining retriever training and the discriminative routing models use a single GPU per run.
\FloatBarrier

\section{Oracle Rule Distribution Analysis}
\label{app:rule-distribution}

\begin{figure}[t]
    \centering
    \includegraphics[width=\textwidth]{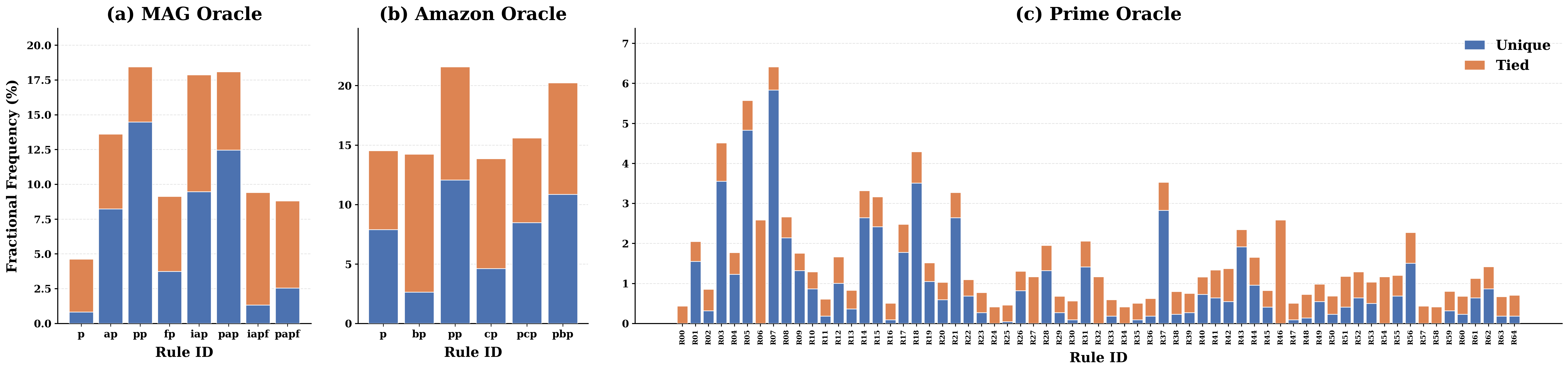}
    \vspace{-3ex}
    \caption{Rule-wise fractional oracle frequency across MAG, Amazon, and PRIME. Each stacked bar decomposes a rule's oracle credit into uniquely optimal (\textbf{Unique}) and tied-optimal (\textbf{Tied}) cases. For a $k$-way tie, each optimal rule receives $1/k$ credit, so the frequencies sum to 100\% within each dataset.}
    \label{fig-rule-dom}
    \vspace{-2ex}
\end{figure}

We analyze how frequently each retrieval rule is optimal under the trained round-1 rule-constraining retrievers. The MAG, Amazon, and PRIME datasets contain 8, 6, and 65 candidate rules, respectively.

For each query $q$ and rule $r$, we compute the mean reciprocal rank over its ground-truth answers $G_q$ as $S(q,r)=\frac{1}{|G_q|}\sum_{a\in G_q}1/\operatorname{rank}_{q,r}(a)$, where an answer absent from the top-200 candidates contributes zero. The oracle set $O_q=\{r:S(q,r)=\max_{r'}S(q,r')\}$ contains all rules achieving the best retrieval score for $q$. A uniquely optimal rule receives full credit, while each rule in a $k$-way tie receives $1/k$ credit. We denote the resulting unique and tied frequencies by $U_r$ and $T_r$, respectively, and define the total fractional oracle frequency as $F_r=U_r+T_r$. Thus, each query contributes total credit one and $\sum_r F_r=1$ within each dataset.

Figure~\ref{fig-rule-dom} reveals three characteristics of the rule space. \textbf{First, oracle utility is highly query-dependent rather than concentrated on a single universally effective rule}. MAG distributes most of its oracle credit among several citation-, authorship-, and institution-based paths, while Amazon exhibits an even broader balance across its smaller rule set. PRIME, despite containing substantially more candidate rules, concentrates useful retrieval behavior in only a subset of them, leaving many rules rarely selected by the oracle. \textbf{Second, a substantial portion of oracle credit arises from ties}, indicating overlap between the retrieval behaviors captured by different rules. This is particularly visible for several MAG and Amazon rules whose oracle contribution is dominated by tied cases. In PRIME, some rules receive oracle credit exclusively through ties: they match the best retrieval quality for certain queries but never provide a uniquely better result. Under the current experts and reward, such rules contribute little distinctive retrieval capability and may therefore be redundant with other rules. \textbf{Finally, the combination of highly uneven usage and strong oracle overlap suggests that the predefined rule space may be over-complete}. Rules with consistent unique oracle credit capture specialized retrieval patterns, whereas rules that are rarely optimal or only tied-optimal are natural candidates for consolidation or pruning. This analysis therefore not only motivates query-specific routing, but also suggests a direction for automatic rule-space refinement, where oracle utility and uniqueness can guide the retention, pruning, or merging of candidate rules.
\FloatBarrier

\section{Case Study}\label{app-case-study}

A central motivation of CoRE is that the reasoning rule for a query is \emph{latent} and cannot be reliably inferred from query semantics alone. When explicit query--rule supervision is unavailable, a natural alternative is to induce rule labels with a strong LLM and train the router on these pseudo-labels. However, such labels are determined only from the query and rule descriptions. The LLM does not observe how each rule-constraining retriever is instantiated, which evidence it can reach, or how well it ranks the ground-truth evidence. As a result, a rule that is semantically plausible may not be the rule that works best for downstream retrieval.

\begin{figure}[t]
    \centering
    \includegraphics[width=\textwidth]{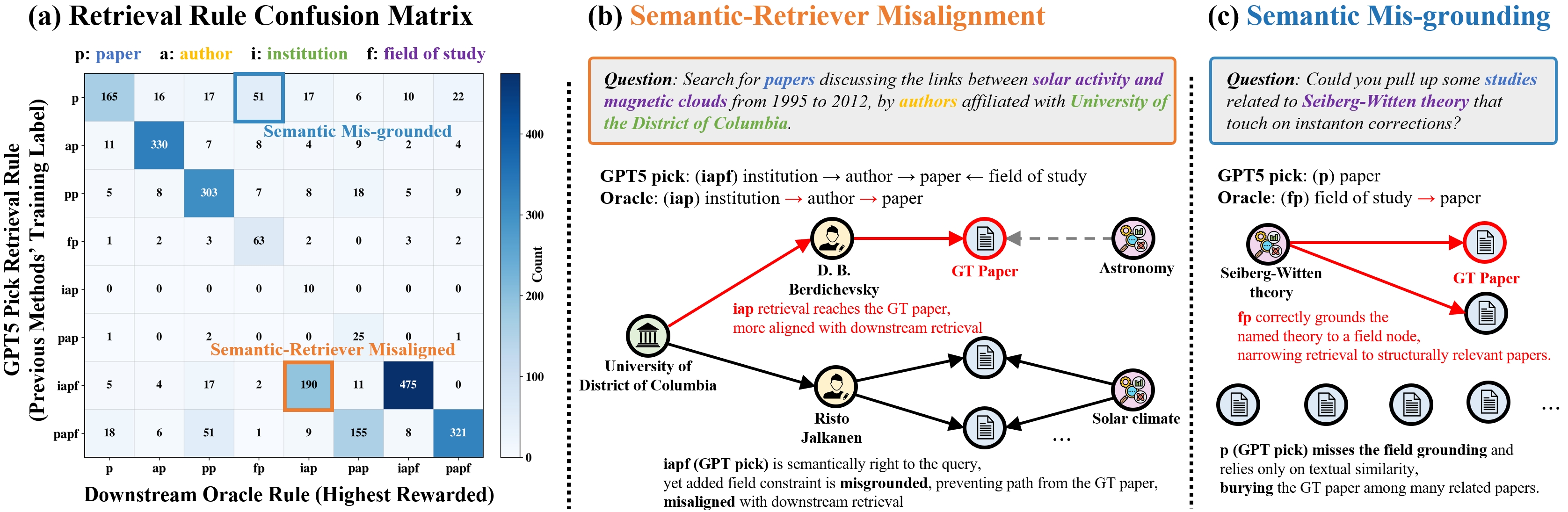}
    \vspace{-3ex}
    \caption{Case study of GPT-5 retrieval-rule routing failures on MAG. (a) \textbf{Confusion matrix} between GPT-5-selected rules and highest-reward oracle rules. (b) {\color[HTML]{ED7D31}\textbf{Semantic--retriever misalignment}}: GPT-5 selects the semantically plausible \texttt{iapf} rule, but the added field constraint is misgrounded by the retriever and prevents access to the ground-truth paper, while the broader \texttt{iap} path succeeds. (c) {\color[HTML]{2272B6}\textbf{Semantic mis-grounding}}: GPT-5 treats the explicit field ``Seiberg--Witten theory'' as free-text content and selects \texttt{p}, whereas \texttt{fp} correctly grounds the field and narrows retrieval to structurally relevant papers.}
    \label{fig-case}
    \vspace{-2ex}
\end{figure}

Figure~\ref{fig-case} examines this gap on MAG using GPT-5 as a strong semantic router. Figure~\ref{fig-case}(a) compares GPT-5-selected rules with the \emph{downstream oracle rule}, defined as the rule whose trained retriever achieves the highest reward for each query. The off-diagonal mass reveals a clear mismatch between semantic routing and retrieval utility. We identify two representative patterns: \textbf{semantic--retriever misalignment}, where the LLM's interpretation is reasonable but the corresponding retriever does not realize it effectively, and \textbf{semantic mis-grounding}, where the router itself maps the query to an inappropriate structural constraint. The former is more common; for example, the \texttt{iapf}$\rightarrow$\texttt{iap} mismatch alone appears in 190 queries.

\textbf{Semantic--retriever misalignment.}
Figure~\ref{fig-case}(b) shows a query containing both an institution and a topical constraint. GPT-5 therefore selects the more specific \texttt{iapf} rule (\emph{institution $\rightarrow$ author $\rightarrow$ paper $\leftarrow$ field of study}), which is semantically reasonable. However, the field branch retrieves field nodes whose paper sets do not overlap with the institution-derived candidates containing the ground-truth papers. Consequently, no valid \texttt{iapf} path reaches the answers. In contrast, the broader \texttt{iap} rule (\emph{institution $\rightarrow$ author $\rightarrow$ paper}) successfully retrieves the two ground-truth papers at ranks 2 and 4. The key issue is therefore not incorrect semantic interpretation, but a mismatch between semantic specificity and actual retriever capability: adding a seemingly relevant constraint can reduce coverage when that constraint cannot be grounded reliably.

\textbf{Semantic mis-grounding.}
Figure~\ref{fig-case}(c) shows the complementary case where the routing decision itself is incorrect. The query explicitly mentions ``Seiberg--Witten theory,'' a valid field-of-study anchor, yet GPT-5 selects direct paper retrieval \texttt{p} and treats the term as ordinary text. Without the field-level structural signal, the ground-truth paper is buried among many topically similar papers and ranks 110th. The \texttt{fp} rule instead grounds the named theory to the corresponding field node and raises the same paper to rank 7. This illustrates a genuine grounding error: the router fails to exploit a structural constraint that is both semantically justified and useful for retrieval.

These cases highlight why static LLM-induced query--rule labels are insufficient for learning a retrieval-oriented router. A good router must capture not only which rule \emph{sounds} appropriate, but which rule works well with the available retrievers and knowledge graph. CoRE addresses this by making routing and retrieval mutually dependent: retriever performance provides the reward for optimizing the routing policy, while updated routing decisions determine the supervision used to specialize the retrievers. The resulting bi-level optimization directly aligns upstream rule selection with downstream retrieval effectiveness.

\FloatBarrier

\section{Different Training Strategies for Classification Router Agents}
\label{app:bert-c-train}

Motivated by the preceding case study, a natural question is whether the router can be trained more simply by treating the highest-reward rule as a pseudo-label. Since classification-based routing allows the retrieval reward of every candidate rule to be evaluated for each query, these rewards can either be converted into supervised targets or optimized directly. We therefore compare three training strategies while keeping the router architecture, retrievers, and co-adaptation procedure unchanged. Let $r_m=1/\mathrm{rank}(S^*\mid Q,K_m)$ denote the reciprocal ground-truth rank under rule $K_m$, and $p_m=P_\theta(K_m\mid Q)$ the router probability. \textbf{(1) Hard-label SFT.} We select the highest-reward rule $m^*=\arg\max_m r_m$ and optimize $\mathcal{L}_{\mathrm{hard}}=-\log p_{m^*}$. This retains only the current best rule and discards reward differences among the remaining rules. \textbf{(2) Soft-label SFT.} We convert the reciprocal-rank rewards into a soft target distribution $q_m=\exp(r_m)/\sum_j\exp(r_j)$ and optimize $\mathcal{L}_{\mathrm{soft}}=-\sum_m q_m\log p_m$, preserving relative rule-level retrieval utility. \textbf{(3) RLFT.} Instead of converting rewards into fixed targets, RLFT directly maximizes expected retrieval utility, using advantages to reinforce above-average rules and suppress lower-reward ones.

\begin{wrapfigure}{r}{0.5\textwidth}
    \vspace{-3ex}
    \centering
    \includegraphics[width=\linewidth]{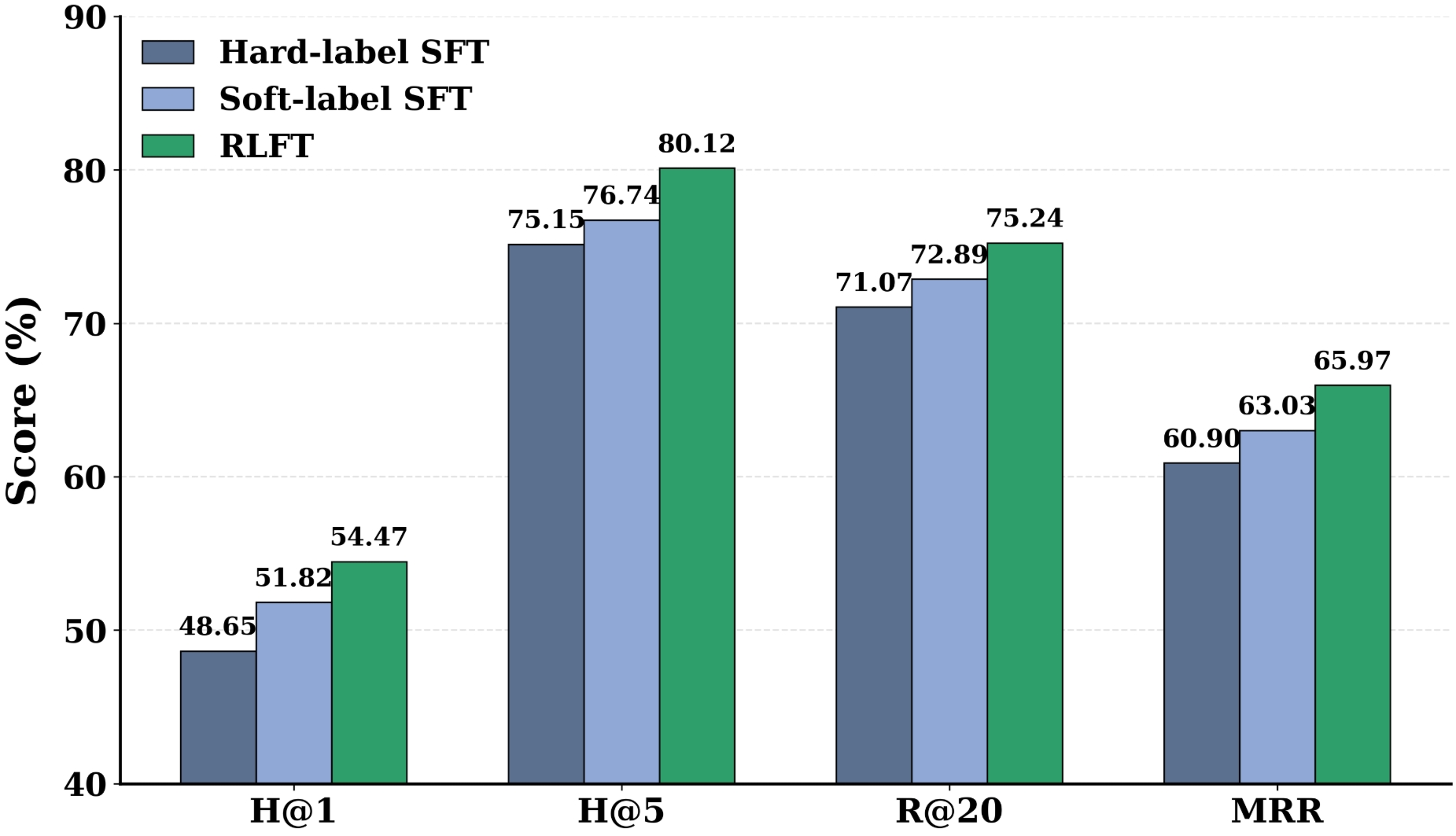}
    \vspace{-3ex}
    \caption{Comparison of training strategies for the BERT-C router, averaged over AMAZON, MAG, and PRIME. RLFT consistently outperforms hard-label and soft-label SFT across all retrieval metrics.}
    \label{fig:c-train}
    \vspace{-2ex}
\end{wrapfigure}
As shown in Figure~\ref{fig:c-train}, performance improves consistently from hard-label SFT to soft-label SFT and further to RLFT across all retrieval metrics. This trend indicates that preserving richer rule-level reward information is more effective than collapsing it into a single pseudo-label, while directly optimizing retrieval utility provides an additional benefit over fitting a reward-derived target distribution.

The difference is also intuitive from their optimization signals. Hard-label SFT only distinguishes the highest-reward rule from all others, regardless of whether its advantage is marginal or substantial. In contrast, RLFT uses the relative retrieval advantage of each rule, producing small updates when rules perform similarly and stronger updates when their retrieval utilities differ substantially. It therefore preserves richer information from the reward landscape instead of reducing it to a binary winner-versus-non-winner signal.

This distinction is particularly important under co-adaptation, where router assignments determine the training data for each rule-constraining retriever. Overly confident early routing can prematurely concentrate supervision on a few rules, limiting the improvement of alternatives as retriever quality adapts. RLFT mitigates this by keeping preferences weak when reward differences are small and strengthening them only as retrieval advantages become clearer.


\newpage
\FloatBarrier

\section{Extended Ablation Study}
\label{sec:extended-ablation}

The main ablation in Table~\ref{tab:ablation} shows that removing either router adaptation or retriever adaptation degrades performance. However, this result alone does not fully distinguish whether the gain comes from \emph{co-adaptation} or simply from using stronger individual components. We therefore conduct two complementary studies that fix one component while varying the other. Together, they separate the effects of routing quality, retrieval quality, and their interaction. Unless otherwise noted, results are averaged over the three STaRK-QA datasets (AMAZON, MAG, and PRIME).

\subsection{Fixing the Retriever, Varying the Router}
\label{subsec:fixed-router}

\begin{wrapfigure}{r}{0.5\textwidth}
    \vspace{-3ex}
    \centering
    \includegraphics[width=\linewidth]{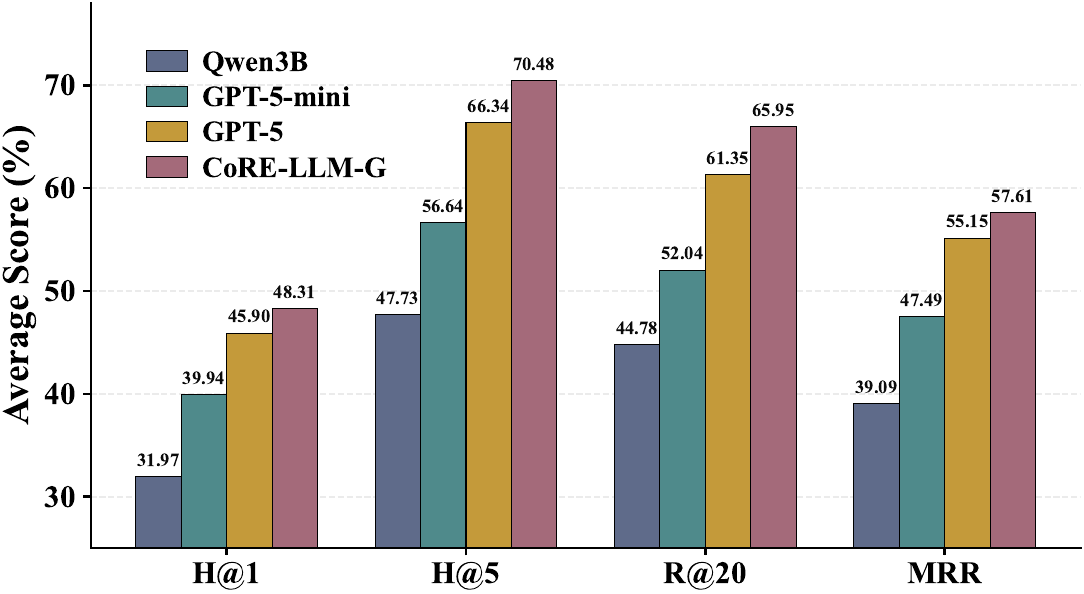}
    \vspace{-3ex}
    \caption{Performance with a fixed trained retriever and different routers, averaged over AMAZON, MAG, and PRIME. Stronger zero-shot LLM routers improve retrieval, but CoRE-LLM-G remains best across all metrics, showing the benefit of retrieval-feedback-based router adaptation.}
    \label{fig:router-ablation}
    \vspace{-4ex}
\end{wrapfigure}

We first isolate the routing effect by fixing the rule-constraining retriever to the one trained under CoRE-LLM-G and replacing only router with three zero-shot general-purpose LLMs: Qwen3B, GPT-5-mini, and GPT-5. We compare these decoupled router substitutions against CoRE-LLM-G, whose router is optimized directly from retrieval feedback.

Figure~\ref{fig:router-ablation} reveals a clear trend. Stronger general-purpose LLMs provide progressively better rule routing, indicating that general reasoning capability transfers to structured rule selection to some extent. However, the improvement saturates before reaching CoRE-LLM-G. Even GPT-5, despite being substantially stronger than the backbone used by CoRE-LLM-G, remains below the co-adaptively trained router on every retrieval metric.
This gap highlights an important distinction between \emph{semantic rule plausibility} and \emph{retrieval-effective rule selection}. A general-purpose LLM can infer which rule appears appropriate from the query semantics, but it does not observe how well each rule is instantiated by the downstream retriever or how effectively that retriever ranks the ground-truth evidence. In contrast, CoRE directly optimizes the routing policy using retrieval outcomes from the paired retriever. As a result, the learned router is calibrated not only to the query, but also to the empirical utility of each rule under the current retrieval system.

Thus, the gain from CoRE cannot be explained by simply replacing the router with a more capable off-the-shelf model. What matters is aligning routing decisions with downstream retrieval behavior through the closed feedback loop.

\vspace{-2ex}
\subsection{Fixing the Router, Varying the Retriever}
\label{subsec:fixed-retriever}

\begin{wrapfigure}{r}{0.4\textwidth}
    \vspace{-3ex}
    \centering
    \includegraphics[width=\linewidth]{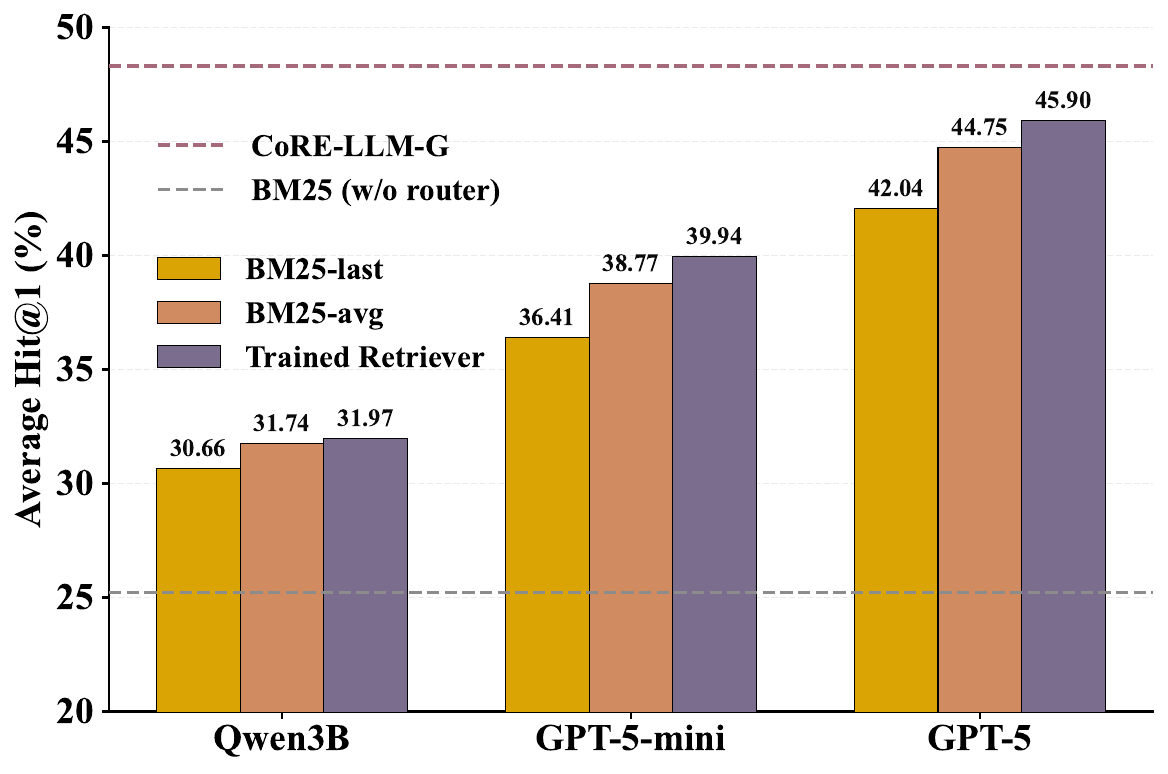}
    \vspace{-3ex}
    \caption{Average Hit@1 across datasets under different router--retriever combinations. Rule-constraining retrieval consistently improves over unconstrained BM25, while stronger retrievers provide additional gains within the selected rule space. CoRE-LLM-G remains best overall.}
    \label{fig:retriever-ablation}
    \vspace{-3ex}
\end{wrapfigure}

We next study the complementary effect of retrieval quality. For each router policy (Qwen3B, GPT-5-mini, and GPT-5), we compare three retrieval strategies under the selected rule constraint: \emph{BM25-last}, which scores candidates using only the terminal entity on the reasoning path; \emph{BM25-avg}, which aggregates BM25 relevance across the full path; and \emph{Trained}, the learnable rule-conditioned retriever used in CoRE. We additionally include plain BM25 over the full knowledge base as an unconstrained reference.
Figure~\ref{fig:retriever-ablation} shows two consistent trends. First, introducing a rule constraint produces the largest qualitative shift in performance. Even simple BM25-based scoring under a routed rule improves over unconstrained BM25, indicating that restricting retrieval to a structurally compatible candidate space is already highly beneficial. This supports the central premise of CoRE: identifying an appropriate reasoning rule is not merely an auxiliary classification problem, but directly determines which evidence can be reached.

Second, once the rule is fixed, retrieval quality still matters. For every router, performance consistently improves from BM25-last to BM25-avg and further to the trained retriever. This suggests that structural restriction and relevance ranking play complementary roles. The router narrows the search space according to the inferred reasoning pattern, while the retriever learns how to rank candidates within that space by exploiting relevance signals along the traversal path.

The gains from retriever improvement are generally smaller than the gains obtained from imposing the rule constraint itself, indicating that routing is the dominant source of improvement over fully unconstrained retrieval. Nevertheless, the consistent benefit of the trained retriever shows that rule selection alone is insufficient: after the correct structural region is identified, the model still benefits from learning how different path-level signals should be combined.

Importantly, the two axes remain complementary rather than substitutable. A stronger router continues to benefit from a stronger retriever, and a stronger retriever continues to benefit from better routing. Yet even the strongest decoupled combination does not match CoRE-LLM-G. This remaining gap is precisely what the co-adaptive objective is designed to address.

Taken together, the two studies clarify the source of CoRE's improvement. Rule routing provides the primary structural advantage by determining the reachable evidence space; retriever adaptation refines ranking within that space; and co-adaptive optimization aligns the two so that routing decisions and retriever specialization reinforce each other. The benefit therefore does not arise from simply scaling either component in isolation.

\FloatBarrier

\section{Inference Time Efficiency Comparison}

\begin{wrapfigure}{r}{0.5\textwidth}
    \vspace{-3ex}
    \centering
    \includegraphics[width=\linewidth]{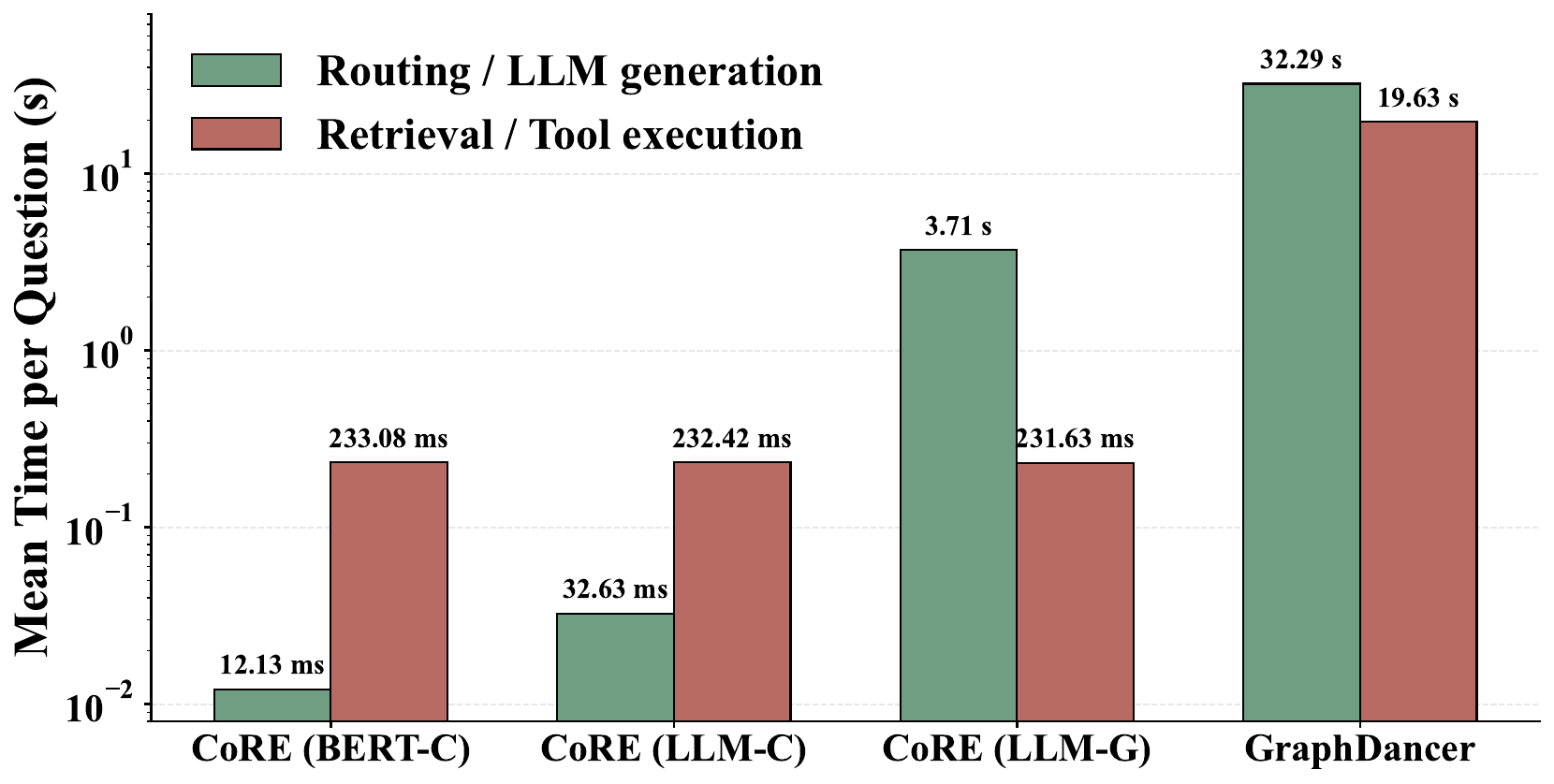}
    \vspace{-3ex}
    \caption{Efficiency comparison on MAG in terms of mean inference time per question. We decompose the cost into routing/LLM generation and retrieval/tool execution. CoRE substantially reduces inference cost compared with step-wise agentic graph exploration, particularly when using classifier-based routing. The y-axis is shown on a logarithmic scale.}
    \label{fig:efficiency}
    \vspace{-2ex}
\end{wrapfigure}

We further examine the inference efficiency of CoRE against the agentic graph reasoning method GraphDancer on MAG. We run each method on 1,000 questions and report the mean inference time per question, decomposed into \textit{routing / LLM generation} and \textit{retrieval / tool execution} costs.

As shown in Figure~\ref{fig:efficiency}, a clear efficiency gap emerges between CoRE and step-wise agentic exploration. The classifier-based variants, CoRE-BERT-C and CoRE-LLM-C, introduce only negligible routing overhead, leaving rule-constraining retrieval as the dominant inference cost. Notably, their retrieval costs are nearly identical, indicating that changing the routing architecture has little effect on the downstream retrieval procedure once a rule has been selected.

CoRE-LLM-G exhibits a different cost profile: its retrieval cost remains comparable to the classifier-based variants, while its routing cost increases substantially due to autoregressive rule generation. This suggests that the additional computational cost of generative routing is concentrated in the routing stage rather than in structured retrieval itself. Nevertheless, its overall inference cost remains substantially below that of GraphDancer.

In contrast, GraphDancer incurs high costs in both components. Its LLM generation cost reflects repeated reasoning over intermediate graph observations, while its tool-execution cost accumulates across multiple exploration steps. The resulting gap highlights a fundamental architectural difference: CoRE performs a single high-level routing decision and delegates subsequent structured search to a specialized rule-constraining retriever, whereas agentic exploration repeatedly interleaves reasoning and graph operations throughout the inference trajectory.

Overall, the results show that rule routing provides a computationally efficient alternative to iterative graph exploration. In particular, classifier-based CoRE achieves strong structured retrieval with minimal decision overhead, while generation-based routing trades additional inference cost for its ability to operate over transferable, tokenized rule representations.

\FloatBarrier

\section{Generation-Based Router Prompt}
\label{app:prompts}


\definecolor{promptback}{RGB}{248,248,248}
\definecolor{promptframe}{RGB}{190,190,190}

\newcommand{\rulekey}[1]{%
    \fcolorbox{gray!35}{gray!8}{%
        \strut\texttt{\footnotesize #1}%
    }%
}

\newcommand{\patharrow}{%
    \allowbreak
    \ensuremath{\mathrel{\;\rightarrow\;}}%
    \allowbreak
}

\newcommand{\corepath}[1]{%
    {\ttfamily #1}%
}

\subsection{Generation Routing Agent Prompts}

The generation-based router selects one rule from a dataset-specific
candidate set. Each input consists of a natural-language query and a set
of reasoning rules represented as schema-level traversal paths, such as
\corepath{Author\patharrow Paper} and
\corepath{Gene/Protein\patharrow Target\patharrow Drug}.
We use the following prompt template across all datasets.

\begin{tcolorbox}[
    enhanced,
    breakable,
    title={Generation-Based Routing Prompt},
    colback=promptback,
    colframe=promptframe,
    coltitle=black,
    fonttitle=\bfseries,
    boxrule=0.6pt,
    arc=2pt,
    left=7pt,
    right=7pt,
    top=6pt,
    bottom=6pt,
    before skip=8pt,
    after skip=10pt
]
\footnotesize

\textbf{System Message}

\medskip

You are a \texttt{\{DOMAIN\}} knowledge graph retrieval rule selector.
Given a query and a set of candidate rules, select the traversal path
that best matches the query.

Consider the target entity, anchor entity, relation semantics, and
traversal direction. Select the simplest rule that fully captures the
query.

\medskip
\textbf{Output Format}

\begin{quote}
\ttfamily
<think>[brief reasoning]</think>\\
<answer>[rule key]</answer>
\end{quote}

\textbf{User Message}

\begin{quote}
\ttfamily
Query: \{QUERY\}

\medskip
Available retrieval rules:\\
\{KEY\_1\}: \{RULE\_1\}\\
\{KEY\_2\}: \{RULE\_2\}\\
\ldots\\
\{KEY\_N\}: \{RULE\_N\}
\end{quote}

Select the single best retrieval rule.

\end{tcolorbox}

Here, \texttt{\{DOMAIN\}} denotes the dataset domain,
\texttt{\{QUERY\}} denotes the input query, and each
\texttt{\{KEY\}: \{RULE\}} pair specifies a valid output key and its
corresponding traversal path. We extract the prediction from the
\texttt{<answer>} field and regard it as valid only if it matches one
of the provided rule keys. Table~\ref{tab:stark-combined} summarizes the
dataset-specific configurations.

\begin{table}[t]
    \centering
    \small
    \caption{Statistics of the STaRK datasets, and generation-based routing configuration.}
    \label{tab:stark-combined}
    \setlength{\tabcolsep}{10pt}
    \renewcommand{\arraystretch}{1.08}
    \begin{tabular}{lccc}
        \toprule
        & \textsc{STaRK-Amazon} & \textsc{STaRK-MAG} & \textsc{STaRK-Prime} \\
        \midrule
        \#entity types    & 4               & 4               & 10 \\
        \#relation types  & 5               & 4               & 18 \\
        \#entities        & 1,035,542       & 1,872,968       & 129,375 \\
        \#relations       & 9,443,802       & 39,802,116      & 8,100,498 \\
        \#tokens          & 592,067,882     & 212,602,571     & 31,844,769 \\
        \midrule
        \#queries         & 9,181           & 13,407          & 11,302 \\
        train             & 5,915           & 7,994           & 6,162 \\
        val               & 1,547           & 2,665           & 2,241 \\
        test              & 1,719           & 2,748           & 2,899 \\
        \midrule
        Rule Keys         & \texttt{A}--\texttt{F} & \texttt{A}--\texttt{H} & \texttt{R00}--\texttt{R64} \\
        \# Rules          & 6               & 8               & 65 \\
        \bottomrule
    \end{tabular}
\end{table}

\subsection{Dataset-Specific Candidate Rules}
\label{app:candidate-rules}

The candidate rule set for each dataset is listed below. Each rule
defines the schema-level traversal used to constrain retrieval.
PRIME contains a substantially larger candidate set. We therefore use
a multi-page table to improve readability. Long traversal paths are
automatically wrapped, preferentially around relation boundaries.




\begin{table}[h]
    \centering
    \small
    \renewcommand{\arraystretch}{1.20}
    \setlength{\tabcolsep}{7pt}
    \caption{Candidate retrieval rules for Amazon.}

    \begin{tabularx}{0.94\linewidth}{
        >{\centering\arraybackslash}p{0.10\linewidth}
        >{\raggedright\arraybackslash}X
    }
        \toprule
        \textbf{Key} & \textbf{Traversal Path} \\
        \midrule

        \rulekey{A}
        & \corepath{Product} \\

        \rulekey{B}
        & \corepath{Brand\patharrow Product} \\

        \rulekey{C}
        & \corepath{Product\patharrow Product} \\

        \rulekey{D}
        & \corepath{Category\patharrow Product} \\

        \rulekey{E}
        & \corepath{Product\patharrow Category\patharrow Product} \\

        \rulekey{F}
        & \corepath{Product\patharrow Brand\patharrow Product} \\

        \bottomrule
    \end{tabularx}
\end{table}



\begin{table}[h]
    \centering
    \small
    \renewcommand{\arraystretch}{1.20}
    \setlength{\tabcolsep}{7pt}
    \caption{Candidate retrieval rules for MAG.}

    \begin{tabularx}{0.94\linewidth}{
        >{\centering\arraybackslash}p{0.10\linewidth}
        >{\raggedright\arraybackslash}X
    }
        \toprule
        \textbf{Key} & \textbf{Traversal Path} \\
        \midrule

        \rulekey{A}
        & \corepath{Paper} \\

        \rulekey{B}
        & \corepath{Author\patharrow Paper} \\

        \rulekey{C}
        & \corepath{Paper\patharrow Paper} \\

        \rulekey{D}
        & \corepath{Field of Study\patharrow Paper} \\

        \rulekey{E}
        & \corepath{Institution\patharrow Author\patharrow Paper} \\

        \rulekey{F}
        & \corepath{Paper\patharrow Author\patharrow Paper} \\

        \rulekey{G}
        & \corepath{
            Institution\patharrow Author\patharrow Paper
            $\leftarrow$ Field of Study
        } \\

        \rulekey{H}
        & \corepath{
            Paper\patharrow Author\patharrow Paper
            $\leftarrow$ Field of Study
        } \\

        \bottomrule
    \end{tabularx}
\end{table}



\small
\renewcommand{\arraystretch}{1.20}
\setlength{\tabcolsep}{7pt}

\begin{longtable}{
    >{\centering\arraybackslash}p{0.10\linewidth}
    >{\raggedright\arraybackslash}p{0.82\linewidth}
}

\caption{Candidate retrieval rules for PRIME.}
\label{tab:prime-rules}\\

\toprule
\textbf{Key} & \textbf{Traversal Path} \\
\midrule
\endfirsthead

\multicolumn{2}{c}{
    \small\itshape
    Table~\ref{tab:prime-rules} continued
}\\
\toprule
\textbf{Key} & \textbf{Traversal Path} \\
\midrule
\endhead

\midrule
\multicolumn{2}{r}{
    \small\itshape Continued on next page
}\\
\endfoot

\bottomrule
\endlastfoot

\rulekey{R00}
& \corepath{
Anatomy\patharrow Expression Present\patharrow Gene/Protein
} \\

\rulekey{R01}
& \corepath{
Anatomy\patharrow Expression Absent\patharrow Gene/Protein
} \\

\rulekey{R02}
& \corepath{
Pathway\patharrow Interacts With\patharrow Gene/Protein
\patharrow Expression Absent\patharrow Anatomy
} \\

\rulekey{R03}
& \corepath{
Gene/Protein\patharrow Target\patharrow Drug
} \\

\rulekey{R04}
& \corepath{
Disease\patharrow Associated With\patharrow Gene/Protein
} \\

\rulekey{R05}
& \corepath{
Gene/Protein\patharrow PPI\patharrow Gene/Protein
} \\

\rulekey{R06}
& \corepath{
Gene/Protein\patharrow Associated With\patharrow Disease
} \\

\rulekey{R07}
& \corepath{
Disease\patharrow Parent-Child\patharrow Disease
} \\

\rulekey{R08}
& \corepath{
Effect/Phenotype\patharrow Phenotype Present\patharrow Disease
} \\

\rulekey{R09}
& \corepath{
Effect/Phenotype\patharrow Phenotype Absent\patharrow Disease
} \\

\rulekey{R10}
& \corepath{
Disease\patharrow Contraindication\patharrow Drug
} \\

\rulekey{R11}
& \corepath{
Gene/Protein\patharrow Expression Absent\patharrow Anatomy
} \\

\rulekey{R12}
& \corepath{
Drug\patharrow Contraindication\patharrow Disease
} \\

\rulekey{R13}
& \corepath{
Biological Process\patharrow Parent-Child\patharrow Biological Process
\patharrow Interacts With\patharrow Gene/Protein
\patharrow Expression Absent\patharrow Anatomy
} \\

\rulekey{R14}
& \corepath{
Drug\patharrow Synergistic Interaction\patharrow Drug
} \\

\rulekey{R15}
& \corepath{
Pathway\patharrow Parent-Child\patharrow Pathway
} \\

\rulekey{R16}
& \corepath{
Drug\patharrow Off-Label Use\patharrow Disease
} \\

\rulekey{R17}
& \corepath{
Molecular Function\patharrow Interacts With\patharrow Gene/Protein
} \\

\rulekey{R18}
& \corepath{
Pathway\patharrow Interacts With\patharrow Gene/Protein
} \\

\rulekey{R19}
& \corepath{
Cellular Component\patharrow Interacts With\patharrow Gene/Protein
} \\

\rulekey{R20}
& \corepath{
Effect/Phenotype\patharrow Associated With\patharrow Gene/Protein
\patharrow Expression Absent\patharrow Anatomy
} \\

\rulekey{R21}
& \corepath{
Disease\patharrow Indication\patharrow Drug
} \\

\rulekey{R22}
& \corepath{
Effect/Phenotype\patharrow Side Effect\patharrow Drug
} \\

\rulekey{R23}
& \corepath{
Exposure\patharrow Linked To\patharrow Disease
} \\

\rulekey{R24}
& \corepath{
Effect/Phenotype\patharrow Associated With\patharrow Gene/Protein
} \\

\rulekey{R25}
& \corepath{
Gene/Protein\patharrow Expression Present\patharrow Anatomy
} \\

\rulekey{R26}
& \corepath{
Drug\patharrow Indication\patharrow Disease
} \\

\rulekey{R27}
& \corepath{
Gene/Protein\patharrow Interacts With\patharrow Pathway
} \\

\rulekey{R28}
& \corepath{
Biological Process\patharrow Interacts With\patharrow Gene/Protein
\patharrow Expression Absent\patharrow Anatomy
} \\

\rulekey{R29}
& \corepath{
Drug\patharrow Synergistic Interaction\patharrow Drug
\patharrow Transporter\patharrow Gene/Protein
} \\

\rulekey{R30}
& \corepath{
Gene/Protein\patharrow Transporter\patharrow Drug
} \\

\rulekey{R31}
& \corepath{
Biological Process\patharrow Interacts With\patharrow Gene/Protein
} \\

\rulekey{R32}
& \corepath{
Gene/Protein\patharrow Interacts With\patharrow Cellular Component
} \\

\rulekey{R33}
& \corepath{
Gene/Protein\patharrow Enzyme\patharrow Drug
} \\

\rulekey{R34}
& \corepath{
Drug\patharrow Enzyme\patharrow Gene/Protein
} \\

\rulekey{R35}
& \corepath{
Disease\patharrow Off-Label Use\patharrow Drug
} \\

\rulekey{R36}
& \corepath{
Drug\patharrow Transporter\patharrow Gene/Protein
\patharrow Interacts With\patharrow Exposure
} \\

\rulekey{R37}
& \corepath{
Drug\patharrow Target\patharrow Gene/Protein
\patharrow Interacts With\patharrow Cellular Component
} \\

\rulekey{R38}
& \corepath{
Disease\patharrow Associated With\patharrow Gene/Protein
\patharrow Expression Absent\patharrow Anatomy
} \\

\rulekey{R39}
& \corepath{
Cellular Component\patharrow Interacts With\patharrow Gene/Protein
\patharrow Expression Absent\patharrow Anatomy
} \\

\rulekey{R40}
& \corepath{
Disease\patharrow Phenotype Present\patharrow Effect/Phenotype
} \\

\rulekey{R41}
& \corepath{
Disease\patharrow Associated With\patharrow Gene/Protein
\patharrow Target\patharrow Drug
} \\

\rulekey{R42}
& \corepath{
Pathway\patharrow Interacts With\patharrow Gene/Protein
\patharrow Interacts With\patharrow Biological Process
} \\

\rulekey{R43}
& \corepath{
Disease\patharrow Parent-Child\patharrow Disease
\patharrow Phenotype Present\patharrow Effect/Phenotype
} \\

\rulekey{R44}
& \corepath{
Pathway\patharrow Interacts With\patharrow Gene/Protein
\patharrow Target\patharrow Drug
} \\

\rulekey{R45}
& \corepath{
Drug\patharrow Transporter\patharrow Gene/Protein
\patharrow Expression Present\patharrow Anatomy
} \\

\rulekey{R46}
& \corepath{
Gene/Protein\patharrow Associated With\patharrow Effect/Phenotype
} \\

\rulekey{R47}
& \corepath{
Gene/Protein\patharrow Carrier\patharrow Drug
} \\

\rulekey{R48}
& \corepath{
Gene/Protein\patharrow PPI\patharrow Gene/Protein
\patharrow Interacts With\patharrow Pathway
} \\

\rulekey{R49}
& \corepath{
Anatomy\patharrow Expression Present\patharrow Gene/Protein
\patharrow Associated With\patharrow Disease
} \\

\rulekey{R50}
& \corepath{
Cellular Component\patharrow Interacts With\patharrow Gene/Protein
\patharrow Target\patharrow Drug
} \\

\rulekey{R51}
& \corepath{
Biological Process\patharrow Interacts With\patharrow Gene/Protein
\patharrow Interacts With\patharrow Pathway
} \\

\rulekey{R52}
& \corepath{
Exposure\patharrow Interacts With\patharrow Gene/Protein
} \\

\rulekey{R53}
& \corepath{
Drug\patharrow Target\patharrow Gene/Protein
} \\

\rulekey{R54}
& \corepath{
Gene/Protein\patharrow Interacts With\patharrow Biological Process
} \\

\rulekey{R55}
& \corepath{
Drug\patharrow Synergistic Interaction\patharrow Drug
\patharrow Target\patharrow Gene/Protein
} \\

\rulekey{R56}
& \corepath{
Molecular Function\patharrow Interacts With\patharrow Gene/Protein
\patharrow Target\patharrow Drug
} \\

\rulekey{R57}
& \corepath{
Biological Process\patharrow Parent-Child\patharrow Biological Process
} \\

\rulekey{R58}
& \corepath{
Gene/Protein\patharrow Transporter\patharrow Drug
\patharrow Side Effect\patharrow Effect/Phenotype
} \\

\rulekey{R59}
& \corepath{
Effect/Phenotype\patharrow Phenotype Present\patharrow Disease
\patharrow Associated With\patharrow Gene/Protein
} \\

\rulekey{R60}
& \corepath{
Molecular Function\patharrow Interacts With\patharrow Gene/Protein
\patharrow Expression Absent\patharrow Anatomy
} \\

\rulekey{R61}
& \corepath{
Anatomy\patharrow Expression Present\patharrow Gene/Protein
\patharrow Enzyme\patharrow Drug
} \\

\rulekey{R62}
& \corepath{
Effect/Phenotype\patharrow Phenotype Present\patharrow Disease
\patharrow Indication\patharrow Drug
} \\

\rulekey{R63}
& \corepath{
Disease\patharrow Associated With\patharrow Gene/Protein
\patharrow Interacts With\patharrow Pathway
} \\

\rulekey{R64}
& \corepath{
Cellular Component\patharrow Interacts With\patharrow Gene/Protein
\patharrow Interacts With\patharrow Pathway
} \\

\end{longtable}

\normalsize

\FloatBarrier





\end{document}